\documentclass[oneside,twocolumn]{article}

\usepackage{geometry}
\usepackage{fullpage}

\usepackage[utf8]{inputenc}
\usepackage{graphicx}
\usepackage{amsmath,mathtools}
\usepackage{amssymb}
\usepackage{amscd}
\usepackage{bm}
\usepackage{url}
\usepackage[super,sort&compress,comma]{natbib}
\usepackage{upgreek}
\usepackage{hyperref}
\usepackage{pdflscape}
\usepackage[font=small,labelfont=bf,labelsep=period]{caption}
\usepackage[version=3]{mhchem}
\usepackage{booktabs}
\usepackage{multirow}
\usepackage{threeparttable}
\usepackage{sectsty}
\usepackage{setspace}
\usepackage{authblk}
\usepackage{fancyhdr}
\usepackage{enumitem}

\usepackage[T1]{fontenc}
\usepackage{textcomp}
\usepackage[euler]{textgreek}

\makeatletter
  \def\@seccntformat#1{\csname the#1\endcsname.\ \ }
\makeatother

\sectionfont{\large}
\subsectionfont{\normalsize}
\subsubsectionfont{\normalsize}

\makeatletter
 \def\@biblabel#1{#1.}
\makeatother
\DeclarePairedDelimiter{\nbr}{(}{)}

\DeclarePairedDelimiter{\sbr}{[}{]}

\newcommand*{\rsub}[1]{\ensuremath{_{\mathrm{#1}}}}
\newcommand*{\rsup}[1]{\ensuremath{^{\mathrm{#1}}}}

\newcommand*{\nga}{\ensuremath{N_{\mathrm{GA}}}}
\newcommand*{\pka}{\ensuremath{\mathrm{p}K_{\mathrm{a}}}}
\newcommand*{\degc}{\ensuremath{^\circ\mathrm{C}}}
\newcommand*{\degree}{\ensuremath{^\circ}}
\newcommand*{\pers}{\ensuremath{\mathrm{s}^{-1}}}

\newcommand*{\us}{{\textmugreek}s}
\newcommand*{\ul}{{\textmugreek}l}
\newcommand*{\hone}{\ce{^{1}H}}
\newcommand*{\htwo}{\ce{^{2}H}}
\newcommand*{\osevt}{\ce{^{17}O}}
\newcommand*{\xe}{\ce{^{129}Xe}}
\newcommand*{\M}{\textsc{m}}
\newcommand*{\mM}{m\textsc{m}}

\makeatletter
\def\@maketitle{%
  \newpage
  \begingroup
  \let \footnote \thanks
  \hrule height \z@
    {\LARGE \bfseries \@title \par}%
    \vskip 6mm
    {\large
      \@author
    }%
  \par\endgroup
  \vskip 5mm 
  \vspace*{0mm}
}
\makeatother

\title{Internal water and microsecond dynamics in myoglobin}
\author{Shuji Kaieda}
\author{Bertil Halle}
\affil{Department of Biophysical Chemistry, Lund University, 
       Lund, Sweden}
\date{}

\begin{document}

%
%
\twocolumn[
\begin{@twocolumnfalse}

\maketitle\thispagestyle{fancy}

%
%
\noindent Myoglobin (Mb) binds diatomic ligands, like \ce{O2}, CO, and NO, in a 
cavity that is only transiently accessible. 
Crystallography and molecular simulations show that the ligands can migrate 
through an extensive network of transiently connected cavities, but disagree on 
the locations and occupancy of internal hydration sites. 
Here, we use water \htwo\ and \osevt\ magnetic relaxation dispersion (MRD) to 
characterize the internal water molecules in Mb under physiological conditions. 
We find that equine carbonmonoxy Mb contains $4.5 \pm 1.0$ ordered internal 
water molecules with a mean survival time of $5.6 \pm 0.5$~\us\ at 25~$\degc$. 
The likely location of these water molecules are the four polar hydration sites, 
including one of the xenon-binding cavities, that are fully occupied in all 
high-resolution crystal structures of equine Mb. 
The finding that water escapes from these sites, located 17 -- 31~\AA\ apart in 
the protein, on the same \us\ time scale suggests a global exchange mechanism. 
We propose that this mechanism involves transient penetration of the protein by 
H-bonded water chains. 
Such a mechanism could play a functional role by eliminating trapped ligands. 
In addition, the MRD results indicate that two or three of the 11 histidine 
residues of equine Mb undergo intramolecular hydrogen exchange on a \us\ time 
scale.
\vspace{0.5cm}
\end{@twocolumnfalse}
]

%
%
\section{\label{sec1}Introduction}

\begin{figure}[!t]
  \centering
  \includegraphics[viewport=190 241 406 601]{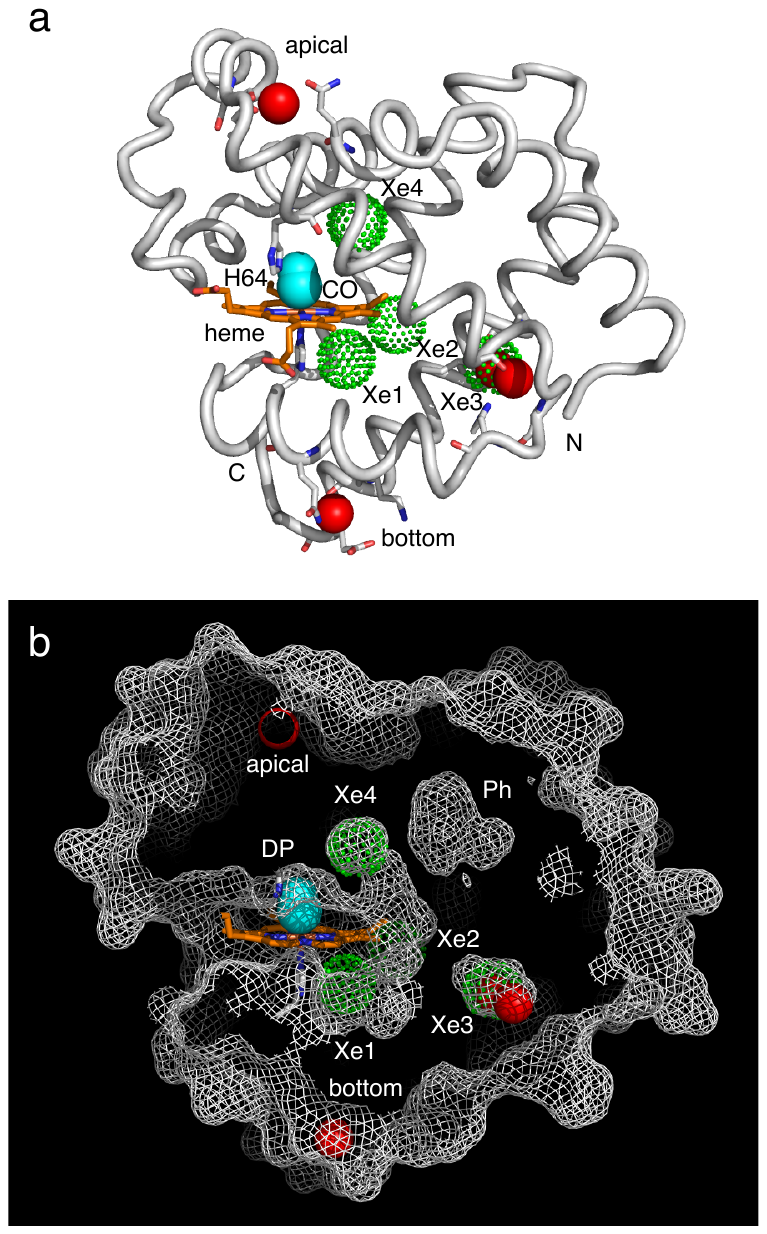}
  \caption{\label{fig1}Crystal structure\cite{Chu2000} 1DWR of equine MbCO in 
           (\textbf{a}) ribbon and (\textbf{b}) cross-sectional surface 
           (1.4~\AA\ probe) representations, rendered with PyMol 
           (Schr\"odinger, LLC). 
           Color code:\ heme (orange), CO (cyan), internal water molecules 
           (red), and xenon binding sites\cite{Tilton1984} (green).}
\end{figure}

The 153-residue hemoprotein myoglobin (Mb) transports \ce{O2} from sarcolemma to 
mitochondria in cardiac and skeletal muscle tissue of most 
mammals.\cite{Wittenberg2003} 
In addition to \ce{O2}, other diatomic ligands, such as CO and NO, bind to the 
heme iron and it has been suggested that Mb also acts as a NO 
scavenger.\cite{Flogel2001} 
Crystal structures\cite{Kendrew1958,Maurus1997,Kachalova1999,Vojtechovsky1999,
Chu2000,Ostermann2002,Hersleth2007} of Mb do not reveal the pathway(s) for 
ligand migration from the exterior to the buried binding site in the distal 
pocket (DP) at the heme-bound iron (Fig.~\ref{fig1}). 
Ligand access to the heme is therefore thought to be controlled by structural 
fluctuations.

Despite a large amount of experimental and computational work, \cite{Austin1975,
Frauenfelder1985,Janes1988,Smith1990,Lim1995a,Olson1996,Brunori2000,Cao2001,
Scott2001,Srajer2001,Schotte2003,Agmon2004,Hummer2004,Nishihara2004,Tetreau2004,
Banushkina2005,Bossa2005,Schmidt2005,Sheu2005,Cohen2006,Goldbeck2006,Anselmi2008,
Ceccarelli2008,Elber2008,Esquerra2008,Ruscio2008,Kiyota2009,Scorciapino2009,
Tomita2009,Cho2010,Elber2010,Esquerra2010,Maragliano2010,Nishihara2010,
Scorciapino2010,Takayanagi2010,Lin2011,Plattner2012,Salter2012,Tsuduki2012,
Boechi2013,Lapelosa2013} the mechanism of conformationally gated ligand 
migration through Mb has not been firmly established. 
In particular, while there is wide support for the direct pathway linking the DP 
to the exterior via the histidine (His-64) gate, the importance of ligand 
migration through the network of cavities in Mb, including the four Xe binding 
sites (Xe1 -- Xe4)\cite{Tilton1984} and the phantom (Ph) region 
(Fig.~\ref{fig1}), is still debated. 
Furthermore, the complex kinetics of ligand migration and escape under 
physiological conditions have not been fully elucidated, although the relevant 
time scale appears to be 0.1 -- 10~\us\ at ambient temperature.

The permanent cavities in Mb, with a combined (probe-size 
dependent\cite{Hubbard1995}) volume approaching 300~\AA$^3$, could accommodate a 
dozen water molecules, but since most of them are lined by nonpolar side-chains 
one expects a low water occupancy.\cite{Rasaiah2008,Yu2010} 
Indeed, high-resolution crystal structures of Mb have identified only four 
(equine Mb)\cite{Maurus1997,Chu2000,Hersleth2007} or five (sperm whale 
Mb)\cite{Kachalova1999,Vojtechovsky1999,Ostermann2002} internal water molecules 
outside the DP. 
The carbonmonoxy form\cite{Kachalova1999,Vojtechovsky1999,Chu2000} of Mb (MbCO) 
has no water in the DP, but a water molecule is found in the DP in the deoxy 
form\cite{Hersleth2007} and is bound to the iron atom in the met form.\cite{
Maurus1997,Kachalova1999,Vojtechovsky1999,Ostermann2002} 
While ligand migration in Mb has been extensively studied, only a few studies 
have examined the involvement of internal water molecules in the mechanism and 
then the focus has been on the DP water.\cite{Cao2001,Goldbeck2006,Esquerra2008,
Esquerra2010}

Information about internal water outside the DP has come from molecular dynamics 
(MD) simulations\cite{Scorciapino2010,Lapelosa2013} and from magnetic relaxation 
dispersion (MRD) experiments.\cite{Aime1993,Zewert1994,Denisov1996a} 
The MD studies indicated several partly occupied internal hydration sites,\cite{
Scorciapino2010,Lapelosa2013} only some of which coincide with the 
crystallographic sites, but these simulations did not access the \us\ time scale 
of water and ligand migration.
 
The water MRD technique measures the magnetic field (or resonance frequency) 
dependence of the longitudinal spin relaxation rate $R_1$ of one of the magnetic 
water nuclides (\hone, \htwo, or \osevt).\cite{Halle1999c,Halle2005} 
Two of the MRD studies\cite{Aime1993,Zewert1994} employed the \hone\ nuclide to 
determine the paramagnetic relaxation enhancement for the DP water in met-Mb. 
The more recent MRD study,\cite{Denisov1996a} which used the \htwo\ and \osevt\ 
nuclides, concluded that equine MbCO contains at least three internal water 
molecules with survival times exceeding the $\sim 10$~ns tumbling time of Mb.

From MRD measurements on protein solutions one can obtain the number of water 
molecules with survival time longer than a lower bound set by the protein 
tumbling time ($\sim 10$~ns for Mb) and shorter than an upper bound set by the 
intrinsic spin relaxation rate for the internal water molecule ($\sim 1$~\us\ 
for \osevt\ and $\sim 100$~\us\ for \htwo).\cite{Halle1999c,Halle2005} 
But within this rather wide range, solution MRD experiments cannot 
differentiate water molecules with different survival times.

More information can be obtained if the protein is immobilized by a 
chemical cross-linking reagent such as glutaraldehyde (GA) so that molecular 
tumbling is inhibited. 
The MRD profile is then essentially the Fourier transform of the survival 
correlation function and survival times up to $\sim 10$~\us\ of individual 
internal water molecules can be determined directly from the \htwo\ MRD 
profile.\cite{Persson2008a,Nilsson2012}

Here, we report an extensive set of \htwo\ and \osevt\ MRD data from equine MbCO. 
Water dynamics on the ns -- \us\ time scale is characterized in detail by 
examining solutions with free-tumbling Mb as well as cross-linked gels with 
immobilized Mb, in both cases under fully hydrated conditions ($\sim 1.6$~\mM\ 
Mb). 
To extract quantitative information about internal water molecules from \htwo\ 
MRD data, it is necessary to isolate any contribution from rapidly exchanging 
labile deuterons in the protein. 
By analyzing the pD dependence of the \htwo\ MRD and by comparing with the 
\osevt\ MRD, we find that several histidine side-chains in Mb exchange deuterons 
with water on the \us\ time scale, apparently by an intramolecular catalytic 
mechanism involving an adjacent deuteron exchange catalyst, such as a 
carboxylate.

The principal questions that we address here are: 
(1) How many internal water molecules occupy the internal cavities of Mb under 
physiological conditions? 
(2) Where are they located? In particular, are the Xe binding sites hydrated? 
(3) How long do the internal water molecules remain inside the cavity network? 
(4) By what mechanism do the internal water molecules exchange? 
(5) Does the internal water play a purely structural role or are they involved 
more directly in the function of Mb?

%
%
\section{\label{sec2}Materials and methods}

%
%
\subsection{\label{sec2.1}Sample preparation}

{\small
Lyophilized equine skeletal muscle Mb was purchased from Sigma ($\geq 95~\%$) 
and further purified by cation-exchange chromatography (SP sepharose; 
GE Healthcare). 
The purified Mb was lyophilized following extensive dialysis against MilliQ 
water. 
To prepare carbonmonoxy Mb (MbCO) samples, the purified protein was dissolved in 
99.8~\% \ce{D2O} (Cambridge Isotope Laboratories) with (gel samples) or without 
(solution samples) a buffer. 
For \osevt\ MRD experiments, 20~\% \osevt-enriched \ce{D2O} was used. 
By using \ce{D2O} for both \htwo\ and \osevt\ MRD experiments, correlation times 
can be compared directly without correcting for the (unknown) H/D kinetic 
isotope effect.

The solution was centrifuged at 13\,000~rpm for 3~min to remove any insoluble 
protein, flushed by CO gas for 3~min on ice, and equilibrated by stirring for 
30~min at 6~$\degc$. 
Mb was then reduced by adding $\sim 2$ equivalents of sodium dithionite (Sigma) 
and the process of flushing CO gas for 2 -- 3~min and stirring for 30~min was 
repeated (usually five times) until no further change in the optical absorption 
spectrum\cite{Antonini1971} (350 -- 650~nm) could be detected (Shimadzu 
UV-1800). 
The Mb concentration was determined from the absorbance at 542~nm with an 
extinction coefficient, 18.4~\mM$^{-1}$\;cm$^{-1}$, that had been calibrated 
using complete amino acid analysis (performed at Amino Acid Analysis Center, 
Dept.\ of Biochemistry and Organic Chemistry, Uppsala University, Sweden). 
MRD measurements were performed on seven solution samples with 2.0 -- 2.6~\mM\ 
Mb and on 26 gel samples, 23 of them with 1.6 -- 1.7~\mM\ Mb 
(Table~\ref{tabS1}).

The protein was immobilized by adding 25~\% GA solution (Sigma) to the MbCO 
solution at 6~$\degc$.\cite{Migneault2004} 
Except for two cases (Table~\ref{tabS1}), the GA/Mb mole ratio, 
$\nga$, was in the range 29.7 -- 31.1. 
After mixing by pipetting, an aliquot ($\sim 50$~\ul) of the solution was 
removed for pH measurement and the remainder was transferred to a glass tube 
(8~mm o.d.\ $\times$ 12~mm height), where it was cross-linked overnight at 
6~$\degc$. 
For the solution samples, pH was adjusted by adding either HCl or NaOH. 
Reported pD values are corrected for the H/D isotope effect according to pD $=$ 
pH$^\ast$ + 0.41, where pH$^\ast$ is the reading on a pH meter calibrated with 
\ce{H2O} buffers.\cite{Covington1968}

To pressurize Xe in a cross-linked MbCO sample, the protein was immobilized in a 
10-mm heavy wall NMR tube with a pressure/vacuum valve (Wilmad). 
After cross-linking, the NMR tube was filled with 8~bar Xe gas (AGA, Sweden) and 
equilibrated for one week before NMR experiments. 
To estimate the Xe concentration in the MbCO sample, two reference samples 
(cyclohexane and 50~\mM\ PIPES pD~7.4 in \ce{D2O}) were prepared. 
The reference solutions were placed in standard 10-mm NMR tubes and Xe gas was 
bubbled through the liquid.
}

%
%
\subsection{\label{sec2.2}NMR experiments}

{\small
The water \htwo\ or \osevt\ longitudinal relaxation rate, $R_1$, was recorded on 
six different NMR set-ups: 
(1) a Stelar Spinmaster 1~T fast field-cycling (FC) spectrometer 
(\htwo:\ 1.5~kHz -- 5.4~MHz); 
(2) a Tecmag Apollo console equipped with a GMW field-variable ($\le 1.8$~T) 
iron-core magnet (\osevt:\ 0.7 -- 1.6~MHz); 
(3) a Tecmag Discovery console equipped with a Drusch field-variable 
($\le 2.1$~T) iron-core magnet (\htwo:\ 2.5 -- 13.1~MHz; \osevt:\ 2.2 -- 
11.5~MHz) or (4) with a Bruker 4.7~T superconducting magnet (\htwo:\ 30.7~MHz; 
\osevt:\ 27.1~MHz); 
and (5) Varian DirectDrive 500 (\osevt:\ 67.8~MHz) and (6) 600 (\htwo:\ 92.1~MHz; 
\osevt:\ 81.3~MHz) spectrometers. 
For the \htwo\ FC measurements, pre-polarized single-pulse ($\leq 2.6$~MHz) or 
inversion recovery ($> 2.6$~MHz) sequences were used with pre-polarization 
and detection at 5.4 and 4.8~MHz, respectively.\cite{Ferrante2005} 
On other spectrometers, the standard inversion recovery pulse sequences were 
used. 
\xe\ spectra were recorded on a Varian DirectDrive 600 (\xe:\ 166~MHz) using a 
single 90$\degree$ pulse. 
All NMR experiments were performed at $25.0 \pm 0.1$~$\degc$, maintained by a 
thermostated air flow. 
The sample temperature was checked with a thermocouple referenced to an 
ice--water bath and by recording the bulk solvent relaxation rate on a reference 
water or buffer sample.
}

%
%
\subsection{\label{sec2.3}Analysis of MRD data}

{\small
Water \htwo\ and \osevt\ MRD profiles, $R_1\nbr{\omega_0}$, from 
\textit{free-tumbling} Mb in \textit{solution} samples were analyzed with the 
standard multi-component exchange model,\cite{Halle1999c,Halle2005} 
\begin{equation}
  R_1\nbr*{\omega_0} = R_1^0 + \frac{1}{N\rsub{W}} 
  \sbr*{N\rsub{H} \nbr*{\xi\rsub{H}-1} R_1^0 +
  \sum_{k} N_k R_1^k\nbr*{\omega_0}}\ ,
  \label{eq1}
\end{equation}
where $\omega_0$ is the Larmor frequency in rad\;$\pers$, $R_1^0$ is the known 
(frequency-independent) bulk solvent relaxation rate and $N\rsub{W}$ is the 
known water/Mb mole ratio in the sample. 
The inverse proportionality of the excess relaxation rate, 
$R_1\rsup{ex}\nbr{\omega_0} \equiv R_1\nbr{\omega_0} - R_1^0$, and $N\rsub{W}$ 
was used to normalize all MRD data to the same Mb concentration, corresponding 
to $N\rsub{W} = 3 \times 10^4$. 
In Eq.~\eqref{eq1}, a component~$k$, as identified by MRD profile analysis, may 
include several sites with similar correlation times (see below).

The first, frequency-independent, term within brackets in Eq.~\eqref{eq1} refers 
to the external hydration shell, comprising $N\rsub{H}$ water molecules with 
average dynamic perturbation factor 
$\xi\rsub{H} \equiv \tau\rsub{H}/\tau_0$.\cite{Mattea2008}  
We estimate $N\rsub{H} = 727$ by computing the solvent-accessible surface area 
for equine Mb (average of PDB structures 1DWR, 1WLA, and 2VIK) with a probe 
radius of 1.7~\AA\ and dividing by the mean surface area, 10.75~\AA$^2$, 
occupied by one water molecule.\cite{Mattea2008} 
The second term within brackets in Eq.~\eqref{eq1} arises from internal water 
molecules (and labile deuterons), with $N_k$ water molecules (or labile-deuteron 
water equivalents) in component~$k$ with intrinsic relaxation rate 
$R_1^k\nbr{\omega_0}$. 
Formally, $N_k$ is the sum of occupancies of the sites belonging to component~$k$. 
$N_k$ can be non-integer since these sites may be partially occupied.

In MbCO, the iron is in the diamagnetic ferrous low-spin state so there is no 
paramagnetic contribution to the water \htwo\ or \osevt\ relaxation rate. 
The intrinsic relaxation rate, induced by the nuclear electric quadrupole 
coupling, is described in the model-free approach as\cite{Halle2009,Halle2005}
\begin{equation}
  R_1^k\nbr*{\omega_0} = S_{\mathrm{iso},k}^{\,2} \omega\rsub{Q}^{\,2} 
  \tau_{\mathrm{C},k} 
  \sbr*{\frac{0.2}{1 + \nbr*{\omega_{0} \tau _{\mathrm{C},k}}^{2}} + 
  \frac{0.8}{1 + \nbr*{2 \omega_{0} \tau_{\mathrm{C},k}}^{2}}}\ .
  \label{eq2}
\end{equation}
The contribution from internal motions in the hydration site, typically on a 
sub-ps time scale, is negligibly small and has been omitted in Eq.~\eqref{eq2}. 
Here, $S_{\mathrm{iso},k}$ is the usual isotropic orientational order 
parameter\cite{Halle1999c,Halle2005,Halle2009} and $\omega\rsub{Q}$ is the 
rigid-lattice nuclear quadrupole frequency ($8.70 \times 10^5$~rad\;$\pers$ for 
\htwo\ and $7.61 \times 10^6$~rad\;$\pers$ for {\osevt}).\cite{Halle1999c} 
The correlation time $\tau_{\mathrm{C},k}$ is related to the tumbling time 
$\tau\rsub{R}$ of the protein (assumed to undergo spherical-top rotational 
diffusion) and the mean survival time (MST) $\tau_{\mathrm{S},k}$ of a water 
molecule in site $k$ as\cite{Halle2009,Halle2005} $1/\tau_{\mathrm{C},k} = 
1/\tau\rsub{R} + 1/\tau_{\mathrm{S},k}$. 
Note that the survival time, often referred to as the residence time in the 
literature, is the time interval from an arbitrary time point to the next water 
exchange event.\cite{Persson2013}

The isotropic rank-2 rotational correlation time $\tau\rsub{R}$ for Mb was 
obtained from molecular hydrodynamics calculations using the program\cite{
Ortega2011} HYDROPRO v.\ 10 with the recommended\cite{Ortega2011,Halle2003b} 
effective (non-hydrogen) atomic radius of 3.0~\AA\ and the 1.45~\AA\ resolution 
crystal structure 1DWR of equine MbCO.\cite{Chu2000} 
Corrected to the viscosity of \ce{D2O} at 25~$\degc$, we thus obtained 
$\tau\rsub{R} = 11.1$~ns, close to the value, $11.9 \pm 0.4$~ns, deduced\cite{
Wang1997} (after correction to \ce{D2O} viscosity) from the field-dependent 
paramagnetic transverse relaxation rate of the proximal histidine 
H$^{\updelta 1}$ in equine deoxy-Mb at 25~$\degc$. 
Because of its oblate-like shape, Mb actually undergoes anisotropic rotational 
diffusion, but the computed 17~\% span of the five asymmetric-top rotational 
correlation times cannot be resolved by our MRD data.

As it stands, Eq.~\eqref{eq1} is strictly valid only in the fast-exchange regime, 
where the MST in each site $k$ that contributes significantly to 
$R_1\nbr{\omega_0}$ is sufficiently short that 
$R_1^k\nbr{0} \tau_{\mathrm{S},k} \ll 1$. 
In the dilute regime, where $N\rsub{W} \gg 1$, however, Eq.~\eqref{eq1} remains 
valid to an excellent approximation for arbitrarily long MST, provided that 
$N_k$ and $\tau_{\mathrm{C},k}$ in Eqs.\ \eqref{eq1} and~\eqref{eq2} are 
reinterpreted as effective parameters according to\cite{Halle2009,Halle2005}
\begin{equation}
  \frac{N_k\rsup{eff}}{N_k} = \frac{\tau_{\mathrm{C},k}\rsup{eff}}
  {\tau_{\mathrm{C},k}} 
  = \sbr*{1 + \omega\rsub{Q}^{\,2} S_{\mathrm{iso},k}^{\,2} 
  \tau_{\mathrm{C},k} \tau_{\mathrm{S},k} }^{-1/2}\ .
  \label{eq3}
\end{equation}

Water \htwo\ MRD profiles from \textit{immobilized} Mb in \textit{gel} samples 
were analyzed with the exchange-mediated orientational randomization (EMOR) 
model,\cite{Nilsson2012} which has recently been quantitatively validated with 
the aid of an ultra-long MD simulation of the protein BPTI.\cite{Persson2013} 
Since the cross-linked protein cannot tumble, the water orientation is 
randomized by the exchange process itself, so that 
$\tau_{\mathrm{C},k} = \tau_{\mathrm{S},k}$. 
As a consequence, the conventional perturbation theory of spin relaxation is 
only valid in the fast-exchange regime, where 
$\omega\rsub{Q} \tau_{\mathrm{S},k} \ll 1$. 
In this regime, the EMOR theory reduces to Eqs.\ \eqref{eq1} and~\eqref{eq2} 
(with $\tau_{\mathrm{C},k} = \tau_{\mathrm{S},k}$). 
Outside this regime, meaning $\tau_{\mathrm{S},k} > 1/\omega\rsub{Q} \approx 
1$~\us, the general form of the EMOR theory, based on the stochastic 
Liouville equation, must be used.\cite{Nilsson2012}
All fits to \htwo\ MRD data from immobilized MbCO reported here are based on 
Eq.~(4.7) of Ref.~\citenum{Nilsson2012}, which is exact in the dilute regime 
($N\rsub{W} \gg 1$). 
Whereas, in the solution case, each component~$k$ is modeled by the two 
parameters $N_k S_{\mathrm{iso},k}^{\,2}$ and $\tau_{\mathrm{C},k}$ (or their 
effective counterparts), there are four parameters in the gel case:\ $N_k$, 
$S_k$, $\eta_k$, and $\tau_{\mathrm{S},k}$.\cite{Nilsson2012} 
Two of these parameters are related to the isotropic order parameter 
as\cite{Nilsson2012} $S_{\mathrm{iso},k} = S_k \nbr{1+\eta_k^{\,2}/3}^{1/2}$.

Since the \osevt\ nuclide has spin quantum number $I = 5/2$, the relaxation 
behavior is in general more complex than for \htwo\ with $I = 1$.\cite{
Halle1999c,Persson2008a} 
In the dilute regime, however, Eqs.\ \eqref{eq1} -- \eqref{eq3} remain valid to 
an excellent approximation in the solution case.\cite{Halle1999c,Halle1981} 
In the gel case, approximate expressions\cite{Persson2008a} are available for 
arbitrary $\tau_{\mathrm{S},k}$, while for $\tau_{\mathrm{S},k} < 
1/\omega\rsub{Q} \approx 100$~ns the EMOR theory reduces to Eqs.\ \eqref{eq1} 
and~\eqref{eq2}, which are valid also for \osevt.\cite{Halle1999c,Halle1981}

In fitting these models to the MRD data we used the trust-region reflective 
nonlinear optimization algorithm\cite{Coleman1996} with uniform 0.5~\% 
(\osevt\ solution MRD) or 1.0~\% (all other MRD profiles) $R_1$ error, estimated 
from the scatter of the frequency independent bulk relaxation rate, $R_1^0$. 
With this algorithm, the model parameters can be constrained to their physically 
admissible ranges.
}

%
%
\section{\label{sec3}Results and discussion}

%
%
\subsection{\label{sec3.1}Overview of MRD}

Because \htwo\ relaxation is relatively slow, the field-cycling technique\cite{
Halle1999c,Ferrante2005} can be used to measure the MRD profile from immobilized 
Mb down to 1.5~kHz, allowing internal-water exchange to be monitored on time 
scales up to 10~\us. 
The drawback is that \htwo\ relaxation does not probe water deuterons 
exclusively, but also may contain a contribution from labile deuterons (LDs) in 
acidic protein side-chains.\cite{Denisov1995b,Halle1999c,VacaChavez2006b,
Persson2008a} 
(This problem is even more severe for \hone\ MRD.\cite{Venu1997,Halle1999c,
VacaChavez2006c}) 
An essential step in the interpretation of \htwo\ MRD data is therefore to 
separate the contributions from \ce{D2O} molecules and LDs. 
To do this, we use four strategies: 
(1) we vary pD to change the acid- and base-catalyzed LD exchange rate and to 
titrate the LD-bearing side-chains; 
(2) we use different buffers to examine buffer-catalyzed LD exchange; 
(3) we compare \htwo\ MRD profiles from immobilized Mb with profiles from 
solutions of free-tumbling Mb, where the LD effect is usually more pronounced; 
and (4) we record \osevt\ MRD profiles (for both immobilized and free-tumbling 
Mb), which monitor water molecules exclusively, but do not extend below 0.7~MHz.

In Secs.\ \ref{sec3.2} -- \ref{sec3.4}, we analyze the \htwo\ and \osevt\ MRD 
data. 
The interpretation of the data gives a consistent picture of the internal 
hydration of Mb, but involves several technical details. 
As a guide to the reader, we summarize the key findings here.

\htwo\ MRD profiles from immobilized MbCO show three dispersion components with 
characteristic mean survival times (MSTs) of 5.6~\us, 121~ns, and 6~ns, 
respectively. 
The one with the slowest MST (5.6~\us) is the dominant component and is 
attributed to internal water molecules and labile \htwo\ in histidine 
side-chains. 
The other two components are assigned to internal motions of His side-chains 
(121~ns) and to confined external water molecules (6~ns) as seen in the 
previous protein gel MRD study.\cite{Persson2008a} 
The pD dependence of the gel \htwo\ MRD profiles demonstrates that $5.2 \pm 0.6$ 
internal water molecules and two or three His side-chains are responsible for 
the 5.6~\us\ component.

Solution \htwo\ MRD profiles from free-tumbling MbCO can be described by a 
single dispersion component. 
This component is attributed to the two slow components (5.6~\us\ and 121~ns) 
and the obtained parameters are consistent with the gel profiles.

An \osevt\ MRD profile from immobilized MbCO reveals an additional internal water 
molecule with an MST of 32~ns. 
This water molecule is not evident in the \htwo\ MRD data, presumably because 
180$\degree$ flips about the water dipole axis reduces the \htwo\ order 
parameter. 
A solution \osevt\ MRD profile is described by a single component just like the 
solution \htwo\ MRD. 
The solution \osevt\ MRD can be accounted for by the 5.6~\us\ and 32~ns 
components, whereas the solution \htwo\ dispersion is produced by the 5.6~\us\ 
and 121~ns components. 
The gel and solution \osevt\ MRD profiles indicate that $3.9 \pm 0.6$ internal 
water molecules are responsible for the 5.6~\us\ component, whereas the \htwo\ 
analysis yields $5.2 \pm 0.6$. 
Our final estimate of $4.5 \pm 1.0$ internal water molecules with an MST of 
5.6~\us\ is the average of the \htwo\ and \osevt\ derived results.

%
%
\subsection{\label{sec3.2}\htwo\ MRD from immobilized MbCO}

\begin{figure}[!t]
  \centering
  \includegraphics[viewport=191 254 404 588]{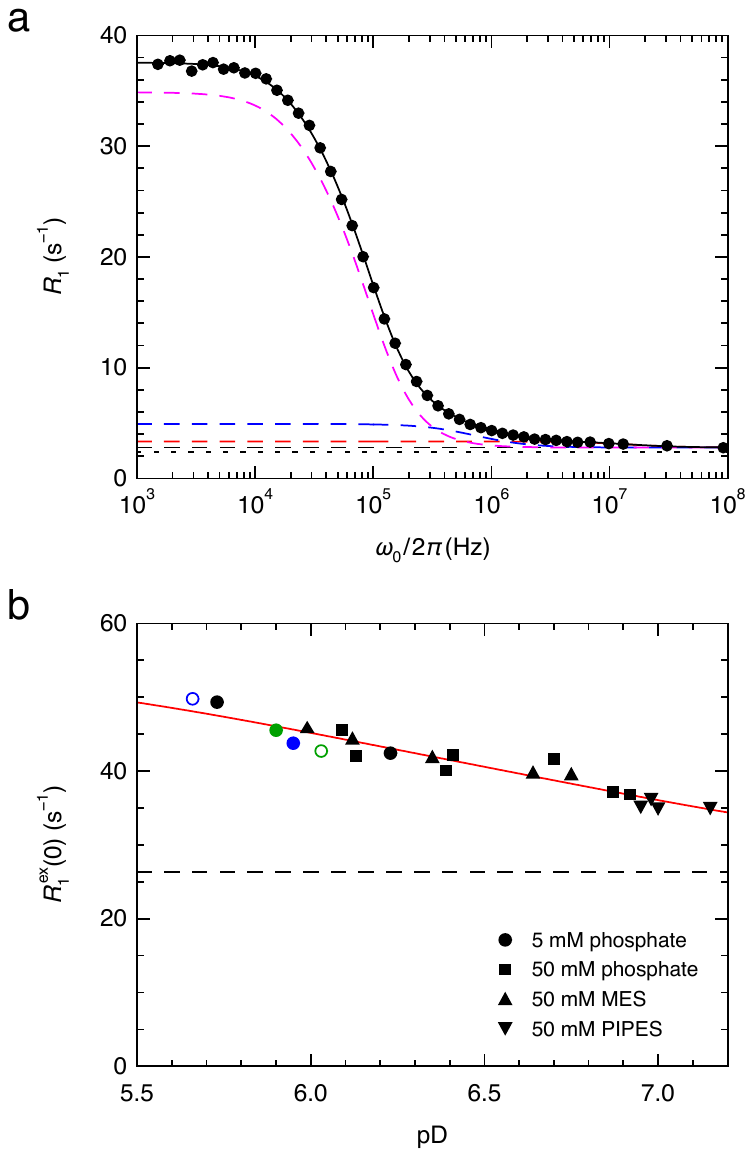}
  \caption{\label{fig2}(\textbf{a}) \htwo\ MRD profile, $R_1\nbr{\omega_0}$, 
           from immobilized MbCO at 25~$\degc$, pD~7.0, and $\nga = 29.9$, 
           scaled to $N\rsub{W} = 3 \times 10^4$. 
           The error bars do not exceed the symbol size. 
           The solid curve is the fit that resulted in the parameter values in 
           Table~\ref{tab1}. 
           The dashed curves are dispersion components 1 (magenta), 2 (blue), 
           and~3 (red). 
           The bulk water (black dotted) and external hydration shell (black 
           dashed) contributions are also shown. 
           (\textbf{b}) Zero-frequency excess relaxation rate, 
           $R_1\rsup{ex}\nbr{0}$, obtained from fits to 22 MRD profiles, versus 
           pD. 
           The red curve is a fit based on the histidine $\pka$ values in Mb and 
           the dashed line is the deduced pD-independent water contribution. 
           The difference between the two curves hence represents the His LD 
           contribution. 
           Solid black symbols represent different buffer conditions as 
           indicated. 
           Colored symbols refer to different MbCO concentrations:\ 2.65~\mM\ 
           (green solid circle), 1.35~\mM\ (green open circle), or 1.02~\mM\ 
           (blue solid circle); and $\nga$:\ 110 (blue solid circle) or 62.1 
           (blue open circle). In all other cases, $C\rsub{Mb} =$ 1.60 -- 
           1.67~\mM\ and $\nga =$ 29.7 -- 31.1.}
\end{figure}

\htwo\ MRD profiles from MbCO immobilized by cross-linking with GA, which reacts 
primarily with lysine side-chains,\cite{Migneault2004} were recorded at 
different Mb concentrations, GA/Mb mole ratios ($\nga$), pD values, buffer 
conditions, and temperatures; in all 25 profiles (Table~\ref{tabS1}). 
Figure~\ref{fig2}a shows a typical profile, measured at pD~7.0. 
The first thing to note is that the low-frequency dispersion has a large 
amplitude, $\sim 3$ times (after $N\rsub{W}$ normalization) that obtained from 
the protein BPTI with three contributing internal water 
molecules.\cite{Persson2008a} 
The dominant contribution to this large dispersion is a component with an MST 
of $5.6 \pm 0.5$~\us\ (Fig.~\ref{fig2}a, Table~\ref{tab1}).
To reveal any LD contribution to this large dispersion, we recorded MRD profiles 
at different pD values in the range 5.66 -- 7.15. 
pD values outside this range were excluded by heme 
dissociation\cite{Antonini1971} and concomitant Mb aggregation (lower pD) or by 
the GA cross-linking chemistry\cite{Migneault2004} (higher pD).

\begin{table}[!t]
  \centering
  \begin{threeparttable}
    \caption{\label{tab1}Results of fits to \htwo\ and \osevt\ MRD 
             profiles.\tnote{a}}
      \footnotesize
      \begin{tabular}{lcccc}
        \toprule
        & \multicolumn{2}{c}{immobilized Mb} 
        & \multicolumn{2}{c}{free-tumbling Mb} \\
        \cmidrule(lr){2-3} \cmidrule(lr){4-5}
         & \htwo & \osevt & \htwo & \osevt \\
        \midrule
        $\tau_{\mathrm{C},1}$ (\us)    & $5.6 \pm 0.5$   & [5.6]         
        & --            & -- \\[1ex]
        $N_1$                          & $6.7 \pm 0.8$   & [5.2]         
        & --            & -- \\[1ex]
        $S_1$                          & $0.79 \pm 0.02$ & [0.79]        
        & --            & -- \\[1ex]
        $\eta_1$                       & $0.3 \pm 0.1$   & [0]           
        & --            & -- \\[1ex]
        $\tau_{\mathrm{C},2}$ (ns)\tnote{b} & $121 \pm 10$    & $32 \pm 3$    
        & $9.5 \pm 0.5$ & $5.6 \pm 0.2$ \\[1ex]
        $N_2 S_{\mathrm{iso},2}^{\,2}$\tnote{b} & $0.70 \pm 0.03$ & $1.0 \pm 0.1$ 
        & $5.6 \pm 0.3$ & $3.0 \pm 0.1$ \\[1ex]
        $\tau_{\mathrm{C},3}$ (ns)     & $6.0 \pm 1.0$   & $4.3 \pm 1.0$ 
        & --            & -- \\[1ex]
        $N_3 S_{\mathrm{iso},3}^{\,2}$ & $3.6 \pm 0.6$   & $4.0 \pm 0.6$ 
        & --            & -- \\[1ex]
        $\xi\rsub{H}$                  & $7.3 \pm 0.5$   & $7.7 \pm 0.4$ 
        & $5.7 \pm 0.5$ & $5.8 \pm 0.1$ \\[1ex]
        $\chi\rsub{red}^{\,2}$         & 1.16            & 0.36          
        & 1.17          & 0.21 \\
        \bottomrule
      \end{tabular}
      \begin{tablenotes}[flushleft]
        \footnotesize
        \item[a] All four MRD profiles were measured at 25~$\degc$ and pD~7.0. 
                 Quoted errors correspond to one standard deviation, propagated 
                 from the uniform 0.5 or 1.0~\% uncertainty in $R_1$. 
                 Parameter values within square brackets were fixed during the 
                 fit.
        \item[b] Component~2 has different physical origins in the four cases 
                 (see text).
      \end{tablenotes}
      \normalsize
  \end{threeparttable}
\end{table}

Figure~\ref{fig2}b shows the pD dependence of the zero-frequency excess 
relaxation rate $R_1\rsup{ex}\nbr{0} \equiv R_1\nbr{0} - R_1^0$, extrapolated 
from the $R_1$ data with the aid of the parameters derived from fits to 22 
individual MRD profiles like the one in Fig.~\ref{fig2}a (Table~\ref{tabS1}). 
Several conclusions can be drawn from these results. 
Since all data points fall on the same master curve despite having been recorded 
on samples with different Mb concentration (1.02 -- 2.65~\mM) and different 
$\nga$ (29.7 -- 110), we infer that the internal-water dynamics probed by the 
MRD profile do not depend significantly on parameters that might affect the 
structure of the cross-linked protein network. 
When the Mb concentration was raised to 5.14~\mM, however, a 23~\% reduction of 
$R_1\rsup{ex}\nbr{0}$ was observed. 
This might be due to impaired exchange for some of internal water molecules or 
labile deuterons caused by increased protein--protein contacts at high 
concentration. 
The subsequent analysis of MRD data from immobilized Mb is restricted to samples 
with $C\rsub{Mb} \approx 1.6$~\mM\ and $\nga \approx 30$.

The second major conclusion that follows from the data in Fig.~\ref{fig2}b is 
that $R_1\rsup{ex}\nbr{0}$ must have a significant LD contribution. 
Before attempting to separate the pD-dependent LD contribution from the 
pD-independent internal-water contribution, we shall identify the side-chains 
and the exchange mechanism responsible for the LD contribution. 
In doing so we are guided by two observational facts: 
(1) the LD contribution to $R_1\rsup{ex}\nbr{0}$ decreases with pD in the 
investigated pD range 5.66 -- 7.15, and (2) the MST, $\tau\rsub{LD}$, of the LDs 
must be close to the 5.6~\us\ MST of the dominant dispersion component 
(Table~\ref{tab1}). 
In the slow-exchange regime ($\tau\rsub{LD} > 1$~\us), 
we expect the LD contribution to $R_1\rsup{ex}\nbr{0}$ to be approximately
proportional to 
$N\rsub{LD}\nbr{\mathrm{pD}}/\tau\rsub{LD}\nbr{\mathrm{pD}}$.\cite{Nilsson2012} 
The observed pD dependence can hence reflect changes in the number, $N\rsub{LD}$, 
of LDs and/or in $\tau\rsub{LD}$.

The LD exchange mechanism can be intermolecular (involving a single side-chain 
and one or more solvent species including buffer ions) or intramolecular 
(involving two proximal side-chains). 
We consider first the usual intermolecular mechanism. 
Due to the observed buffer independence of $R_1\rsup{ex}\nbr{0}$ 
(Fig.~\ref{fig2}b), buffer catalyzed LD exchange can be neglected. 
We therefore need to consider only the acid (\ce{D3O+}) and base (\ce{OD-}) 
catalyzed exchange mechanisms. 
LD exchange from basic groups (Arg, Lys, His, N-terminus) is base catalyzed in 
the examined pD range. 
Any LD contribution from these groups would vary with pD as 
$N\rsub{LD}/\tau\rsub{LD} \propto 1/\nbr{10^{-\pka} + 10^{-\mathrm{pD}}}$, 
that is, it would increase monotonically with pD, in contrast to the observed 
decrease (Fig.~\ref{fig2}b). 
Furthermore, even if base catalyzed LD exchange proceeds at the maximum 
(diffusion controlled) rate, $\tau\rsub{LD}$ at pD 7 is three orders of 
magnitude longer than the required \us\ time scale. 
LD exchange from hydroxyl-bearing groups (Ser, Thr, Tyr) is both acid 
and base catalyzed. 
In principle, the observed pD dependence (Fig.~\ref{fig2}b) could therefore 
reflect acid catalysis. 
The rate constant, $k\rsub{a}$, for acid catalysis is, however, typically four 
orders of magnitude smaller than the rate constant, $k\rsub{b}$, for base 
catalysis.\cite{Denisov1995b,VacaChavez2006b,Liepinsh1996} 
The minimum LD exchange rate should therefore occur close to 
$\mathrm{pD}^\ast = \nbr{1/2} \log\sbr{k\rsub{a}/\nbr{K\rsub{W} k\rsub{b}}} 
\approx 5.5$, where $K\rsub{W} = 10^{-14.95}$. 
In the examined pD range, we would therefore expect $R_1\rsup{ex}\nbr{0}$ to 
increase with pD if hydroxyl LDs were responsible. 
Furthermore, even with the maximum conceivable rate constants\cite{Denisov1995b,
VacaChavez2006b,Liepinsh1996} ($k\rsub{a} \approx 10^7$~\M$^{-1}\;\pers$ and 
$k\rsub{b} \approx 10^{11}$~\M$^{-1}\;\pers$), $\tau\rsub{LD}$ is too long, by 
at least two orders of magnitude, in the examined pD range. 
LD exchange from carboxyl groups (Asp, Glu, C-terminus) usually occurs directly 
to water without involvement of catalytic species, so that $\tau\rsub{LD}$ is 
independent of pD.\cite{Denisov1995b,VacaChavez2006b} 
In principle, the observed pD dependence (Fig.~\ref{fig2}b) could therefore 
reflect deprotonation of COOD groups. 
For immobilized BPTI, a small COOD contribution was seen at pD 4.4, but not at 
pD 6.5 where all carboxyl groups in BPTI are deprotonated.\cite{Persson2008a} 
For MbCO, carboxyl groups with strongly upshifted $\pka$ are not 
indicated,\cite{Bashford1993} so carboxyl LDs cannot explain the pD dependence 
in Fig.~\ref{fig2}b.

Based on these considerations, we conclude that the LD contribution observed 
here is produced by intramolecular LD exchange with an essentially pD-independent 
rate. 
The observed pD dependence in $R_1\rsup{ex}\nbr{0}$ must then reflect the ionization 
equilibrium. 
With no $\pka$ upshifted carboxyl group, only the imidazolium (His) groups are 
titrated in the examined pD range. 
Equine Mb contains 11 histidines (BPTI has none) with $\pka$ values in the 
neutral range (Table~\ref{tabS2}).\cite{Bashford1993,Bhattacharya1997,Kao2000,
Rabenstein2001} 
We then expect that $R_1\rsup{ex}\nbr{0} = a + N\rsub{LD}\nbr{\mathrm{pD}} b$, 
where $a$ is the internal water contribution and $N\rsub{LD}$ is the number of 
labile His deuterons (two in the acidic form, none in the basic form). 
Using published $\pka$ values,\cite{Bashford1993,Bhattacharya1997,Kao2000,
Rabenstein2001} we find that this function can reproduce the observed pD 
dependence in $R_1\rsup{ex}\nbr{0}$. 
The fit shown in Fig.~\ref{fig2}b yields a pD-independent contribution (the 
parameter $a$) to $R_1\rsup{ex}\nbr{0}$ of 26.4~$\pers$, corresponding to 73.1~\% of 
$R_1\rsup{ex}\nbr{0}$ at pD 7.0. 
Although all the His residues were included in this analysis, the result only 
depends on the $\pka$ distribution (in the examined pD range) and not on the 
number of His residues. 
Indeed, very similar results are obtained (26.2~$\pers$ and 73.3~\%) if we 
assume that the pD dependence is produced by a single His residue with a typical 
$\pka$ value of 6.5.

The LDs of His side-chains can exchange via an intramolecular exchange mechanism, 
where a water molecule bridges a positive His residue and negative carboxylate 
group or a positive/neutral His pair (Sec.~\ref{secS1}), 
similar to what has been proposed for certain low-molecular-weight His 
derivatives.\cite{Ralph1980} 
The crystal structure of equine MbCO reveals several His residues that could 
engage in such intramolecular exchange (Sec.~\ref{secS1}, Fig.~\ref{figS1}).
A more detailed discussion of different exchange mechanisms for His LDs can be 
found in Sec.~\ref{secS1}.

The \htwo\ MRD profiles from immobilized MbCO are strongly dominated by the 
slowest kinetic component (Fig.~\ref{fig2}a), hereafter referred to as 
component~1. 
At pD~7.0, the fit yields $N_1 = 6.7 \pm 0.8$ for the number of water molecules 
and/or LD water equivalents (one acidic His residue equates to one \ce{D2O} 
molecule) responsible for this component (Table~\ref{tab1}). 
Hydration sites may have fractional occupancy and His residues may be partly 
titrated, so $N_1$ need not be an integer. 
The analysis in Fig.~\ref{fig2}b indicates that 73.1~\% of $R_1\rsup{ex}\nbr{0}$ 
is produced by water deuterons at pD 7.0 and the MRD fit in Fig.~\ref{fig2}a 
shows that $R_1\rsup{ex}\nbr{0}$ at pD~7.0 can be broken down as follows: 
91.3~\% component~1 (with contributions from water and LDs), 6.1~\% component~2 
(entirely due to LDs; see below), and 2.6~\% from component~3 and the 
frequency-independent hydration-shell component (both entirely due to water; see 
below). 
We thus infer that $(73.1 - 2.6)/0.913 = 77.2$~\% of component~1 is produced by 
$N_1\rsup{W} = 0.772 N_1 = 5.2 \pm 0.6$ internal water molecules.

The remaining 22.8~\% of component~1 at pD~7.0 is attributed to LDs in His 
residues (see above). 
If the locally averaged nuclear quadrupole coupling is the same for the LDs as 
for the $\sim 5$ internal water molecules, we can infer that $N_1\rsup{LD} = 3.0 
\pm 0.4$ LDs contribute to component~1 at pD~7.0. 
Since the (rigid-lattice) \htwo\ quadrupole coupling constant for the acidic 
imidazole deuterons\cite{Zhao2006} is expected to be smaller than for \ce{D2O}, 
the actual value of $N_1\rsup{LD}$ may be as large as five. 
We therefore conclude that at least two, and possibly three, His side-chains 
(with two LDs per imidazolium group) engage in water-mediated intramolecular LD 
exchange on the \us\ time scale.

According to the fit, the MST of the water molecules contributing to component~1 
is $\tau\rsub{S,1} = 5.6 \pm 0.5$~\us\ (Table~\ref{tab1}). 
That $\sim 5$ internal water molecules, located at different sites within the 
MbCO molecule, have the same MST suggests that they exchange by a common 
mechanism (Sec.~\ref{sec3.6}). 
But it should be noted that even though the dominant low-frequency dispersion is 
reproduced by a single kinetic EMOR component, a modest MST distribution would 
not have been resolved by our data.

The value $\tau\rsub{S,1} = 5.6$~\us\ is on the long side of the maximum of 
$R_1\nbr{0}$ versus $\tau\rsub{S}$, corresponding to $\tau\rsub{S} \approx 
1/\omega\rsub{Q} \approx 1$~\us.\cite{Persson2008a,Nilsson2012} 
In this slow-exchange regime, $R_1\nbr{0}$ decreases with increasing 
$\tau\rsub{S}$. 
For example, if $\tau\rsub{S,1}$ had been 1~\us, $R_1\nbr{0}$ would have been 
several-fold larger and the dispersion would have appeared at a higher 
frequency. 
If $\tau\rsub{S,1}$ had been much longer, say 50~\us, then the dispersion 
frequency would only have been slightly downshifted, but the amplitude would 
have been strongly reduced.\cite{Nilsson2012} 
In other words, a much larger number of internal water molecules or LDs would 
have been required to account for the observed MRD profile if the MST were much 
longer. 
In fact, in the ultraslow motion regime, where 
$\tau\rsub{S} \gg 1/\omega\rsub{Q} \approx 1$~\us, $R_1\nbr{\omega_0}$ is 
approximately proportional to $N/\tau\rsub{S}$.\cite{Nilsson2012} 
Even though the observed 5.6~\us\ MST is not long enough to be in this regime, 
the parameters $N_1$ and $\tau\rsub{S,1}$ have a sizeable covariance. 
Reassuringly, constrained fits to the \htwo\ MRD profile with different fixed 
values of $N_1$ yield the best fit quality for $N_1$ between six and seven 
(Fig.~\ref{figS2}), consistent with the result, $N_1 = 6.7 \pm 0.8$, from the 
unconstrained fit.

To reproduce the \htwo\ MRD profile also at higher frequencies, the model must 
include, in addition to the dominant 5~\us\ component, two components in the 
fast-exchange regime (Table~\ref{tab1}). 
One of these components has $\tau\rsub{S,2} = 0.12 \pm 0.01$~\us\ and 
$N_2 S\rsub{iso,2}^{\,2} = 0.70 \pm 0.03$. 
As argued in Sec.~\ref{sec3.4}, this component can be attributed to internal 
motions of rapidly exchanging LDs in His side-chains. 
The third dispersion component has $\tau\rsub{S,3} = 6 \pm 1$~ns and 
$N_3 S\rsub{iso,3}^{\,2} = 3.6 \pm 0.6$ (Table~\ref{tab1}). 
Similar parameter values were obtained for the highest-frequency component of 
the \htwo\ MRD profiles of immobilized BPTI and ubiquitin.\cite{Persson2008a} 
As before, we attribute this component to a small number of confined water 
molecules in the external hydration shell.\cite{Persson2008a} 
Finally, the dynamic perturbation factor, $\xi\rsub{H} = 7.3 \pm 0.5$, deduced 
from the frequency-independent excess relaxation, is somewhat larger than for 
free-tumbling Mb (Sec.~\ref{sec3.3}). 
This was also the case for immobilized BPTI ($\xi\rsub{H} = 7.9 \pm 0.2$) and 
ubiquitin ($6.0 \pm 0.2$),\cite{Persson2008a} both of which yield $\xi\rsub{H} 
\approx 4$ for the free-tumbling proteins.\cite{Mattea2008} 
This difference between gel and solution samples, as well as the fact that 
component~3 is not observed in solution (Sec.~\ref{sec3.3} and \ref{sec3.4}), 
can be rationalized in terms of gel-induced confinement due to short 
intermolecular cross-links and/or intramolecular cross-links (equine Mb has 19 
Lys residues).

In summary, the high-frequency tail of the \htwo\ MRD profile from immobilized 
MbCO exhibits the expected features associated with the external hydration 
shell, whereas the low-frequency part (components 1 and~2) reflects the 
occupancy and dynamics of internal hydration sites and LDs in a few His 
residues. 
The observed pD dependence of the \htwo\ MRD profile indicates that 
$5.2 \pm 0.6$ internal water molecules with MST $5.6 \pm 0.5$~\us\ contribute to 
the dominant dispersion component~1.

%
%
\subsection{\label{sec3.3}\htwo\ MRD from free-tumbling MbCO}

As a check on the interpretation of the \htwo\ MRD data from immobilized MbCO, 
we measured the \htwo\ MRD profile from free-tumbling MbCO in solutions with pD 
in the range 5.0 -- 10.0. 
(For the gel samples, the cross-linking chemistry\cite{Migneault2004} limited us 
to pD~$\le 7.2$.) 
In the acidic pD range, a low-frequency MRD component was observed with a 
correlation time corresponding to tumbling of large protein aggregates, 
presumably induced by heme dissociation.\cite{Antonini1971} 
At pD~5.0, this component dominated the \htwo\ MRD profile. 
We therefore restrict the analysis to $R_1$ data measured at 
$\omega_0/2\pi > 0.9$~MHz 
on solutions with pD~$\ge 6$. 
Under these conditions, the fraction aggregated protein is so small ($\ll 1$~\%) 
that the associated $R_1$ contribution can be neglected.

\begin{figure}[!t]
  \centering
  \includegraphics[viewport=190 254 405 588]{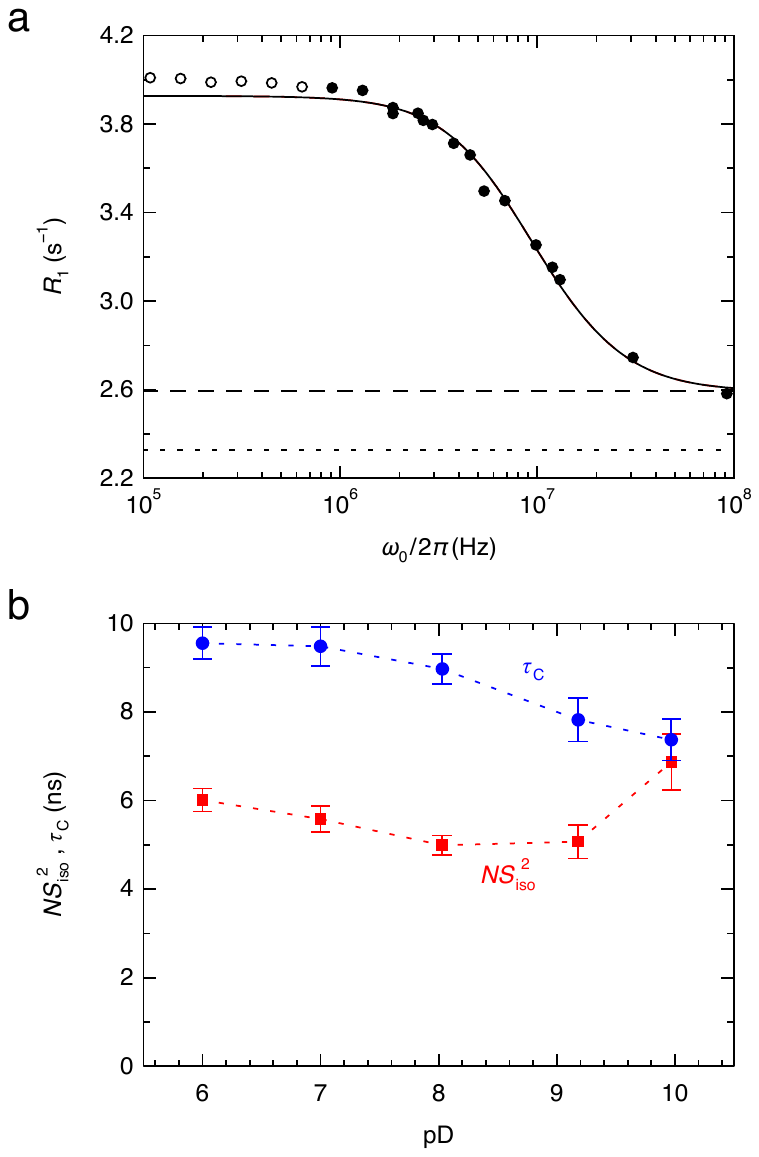}
  \caption{\label{fig3}(\textbf{a}) \htwo\ MRD profile, $R_1\nbr{\omega_0}$, 
           from free-tumbling MbCO at 25~$\degc$ and pD~7.0, scaled to 
           $N\rsub{W} = 3 \times 10^4$. 
           The error bars are up to twice the symbol size. 
           The solid curve is a single-component fit to the solid data points 
           ($\omega_0/2\pi > 0.9$~MHz), with the parameters in Table~\ref{tab1}. 
           The bulk water (dotted line) and external hydration shell (dashed 
           line) contributions are shown. 
           (\textbf{b}) Amplitude $N S\rsub{iso}^{\,2}$ (red squares) and 
           correlation time $\tau\rsub{C}$ (blue circles), obtained from 
           single-component fits to five MRD profiles, versus pD. 
           The dotted lines are visual guides.}
\end{figure}

Figure~\ref{fig3}a shows the \htwo\ MRD profile from free-tumbling MbCO at 
pD~7.0. 
At all pD values in the range 6.0 -- 10.0, the MRD data (for 
$\omega_0/2\pi > 0.9$~MHz) can be adequately represented by a single dispersion 
component. 
The correlation time, $\tau\rsub{C}$, deduced from the single-component fit is 
shorter than the Mb tumbling time, $\tau\rsub{R} = 11.1$~ns (Sec.~\ref{sec2.3}) 
and it decreases with increasing pD: from $9.6 \pm 0.4$~ns at pD~6.0 to 
$7.4 \pm 0.5$~ns at pD~10.0 (Fig.~\ref{fig3}b). 
These findings can be explained if there is a pD-dependent contribution from LDs 
exchanging on the $\sim 100$~\us\ time scale, as expected for base-catalyzed LD 
exchange in hydroxyl, ammonium, and guanidinium groups.\cite{Persson2008a,
Denisov1995b,VacaChavez2006b,Liepinsh1996} 
In contrast to the rapidly exchanging His LDs (Sec.~\ref{sec3.2}), these more 
slowly exchanging LDs do not contribute to the \htwo\ MRD profile from 
immobilized MbCO because of the more restrictive fast-exchange criterion 
$\omega\rsub{Q} \tau\rsub{S} < 1$. 
For a free-tumbling protein, the fast-exchange criterion is 
$\omega\rsub{Q} \nbr{\tau\rsub{R} \tau\rsub{S}}^{1/2} < 1$ (Sec.~\ref{sec2.3}), 
so LDs with $\tau\rsub{S}$ on the order of 100~\us\ make a significant 
contribution, albeit with reduced effective weight $N\rsup{eff}$ and correlation 
time $\tau\rsub{C}\rsup{eff}$ according to Eq.~\eqref{eq3}. 
As explained in more detail in Sec.~\ref{secS2}, $\tau\rsub{C}$ decreases with pD 
(Fig.~\ref{fig3}b) because the relative contribution to the MRD profile from LDs 
with reduced effective correlation time increases with pD.

The pD dependence of the amplitude parameter $N S\rsub{iso}^{\,2}$ has a broad 
minimum of $5.0 \pm 0.3$ at pD 8 -- 9 (Fig.~\ref{fig3}b). 
It decreases first because the His residues involved in intramolecular LD 
exchange are titrated (Sec.~\ref{sec3.2}), consistent with the decrease of 
$R_1\rsup{ex}\nbr{0}$ with pD in Fig.~\ref{fig2}b. 
Above pD~$\sim 9$, where all His residues in equine Mb are titrated,\cite{
Bashford1993,Bhattacharya1997,Kao2000,Rabenstein2001} 
$N S\rsub{iso}^{\,2}$ increases because base-catalyzed exchange brings an 
increasing number of hydroxyl-bearing or basic (mainly Arg) side-chains in Mb 
into the fast-exchange regime.\cite{Denisov1995b}

The dynamic perturbation factor $\xi\rsub{H}$ for the external hydration shell 
does not vary significantly in the pD range 6.0 -- 8.0, where $\xi\rsub{H} = 
5.5 \pm 0.2$ (mean $\pm$ standard deviation). 
This value is within the range found for other proteins.\cite{Mattea2008,
Halle2004a} 
At pD~10, $\xi\rsub{H} = 6.5 \pm 0.6$ is slightly larger, most likely due to a 
frequency-independent $R_1$ contribution from sub-ns internal motions of rapidly 
exchanging LDs.\cite{Denisov1995b}

The single component in the solution MRD profile (Fig.~\ref{fig3}a) must include 
both components 1 and~2 in the gel MRD profile (Fig.~\ref{fig2}a). 
In solution at pD~7.0, these two components should appear as a single 
(unresolved) component with amplitude (see Sec.~\ref{sec2.3}) 
$N S\rsub{iso}^{\,2} = N_1 S_1^{\,2} \nbr{1+\eta_1^{\,2}/3} 
+ N_2 S\rsub{iso,2}^{\,2} = 5.0 \pm 0.9$ (Table~\ref{tab1}), consistent with the 
value $N S\rsub{iso}^{\,2} = 5.6 \pm 0.3$ deduced from the solution profile in 
Fig.~\ref{fig3}a. 
A slightly different way of showing this consistency is to compute 
$N S\rsub{iso}^{\,2}$ as a function of $N_1$ using parameters from constrained 
gel MRD fits with different fixed $N_1$ values. 
The intersection of the resulting $N S\rsub{iso}^{\,2} = f\nbr{N_1}$ curve with 
the solution value $5.6 \pm 0.3$ yields $N_1 = 7.2 \pm 0.6$ (Fig.~\ref{figS3}). 
In this analysis, we ignored the minor contribution from slowly exchanging LDs 
to the solution $N S\rsub{iso}^{\,2}$ at pD~7.0. 
Using the minimum (at pD~8.0) solution value $N S\rsub{iso}^{\,2} = 5.0 \pm 0.2$ 
we obtain instead $N_1 = 6.6 \pm 0.6$. 
In either case, the result of this analysis provides further support for the 
value $N_1 = 6.7 \pm 0.8$ obtained from the unconstrained fit (Table~\ref{tab1}).

In summary, the mutually consistent \htwo\ MRD profiles from immobilized and 
free-tumbling protein indicate that MbCO contains $5.2 \pm 0.6$ highly ordered 
($S = 0.79 \pm 0.02$) internal water molecules with a common MST of 
$5.6 \pm 0.5$~\us. 
In addition, the MRD profiles contain pD-dependent contributions from LD 
exchange and internal motions in His (neutral pD) and other (basic pD) 
side-chains.

%
%
\subsection{\label{sec3.4}\osevt\ MRD from immobilized and free-iumbling MbCO}

While \osevt\ MRD is rigorously water selective, the fast \osevt\ relaxation 
precludes the use of field-cycling to access the low-frequency range of the 
5~\us\ dispersion.\cite{Halle1999c} 
But even if this were possible, a water molecule with MST of 5~\us\ would be far 
into the slow-exchange regime (since $\omega\rsub{Q}^{\,2}$ is two orders of 
magnitude larger for \osevt\ than for \htwo) and it would therefore contribute 
very little to the \osevt\ MRD profile from immobilized MbCO.

\begin{figure}[!t]
  \centering
  \includegraphics[viewport=189 253 406 589]{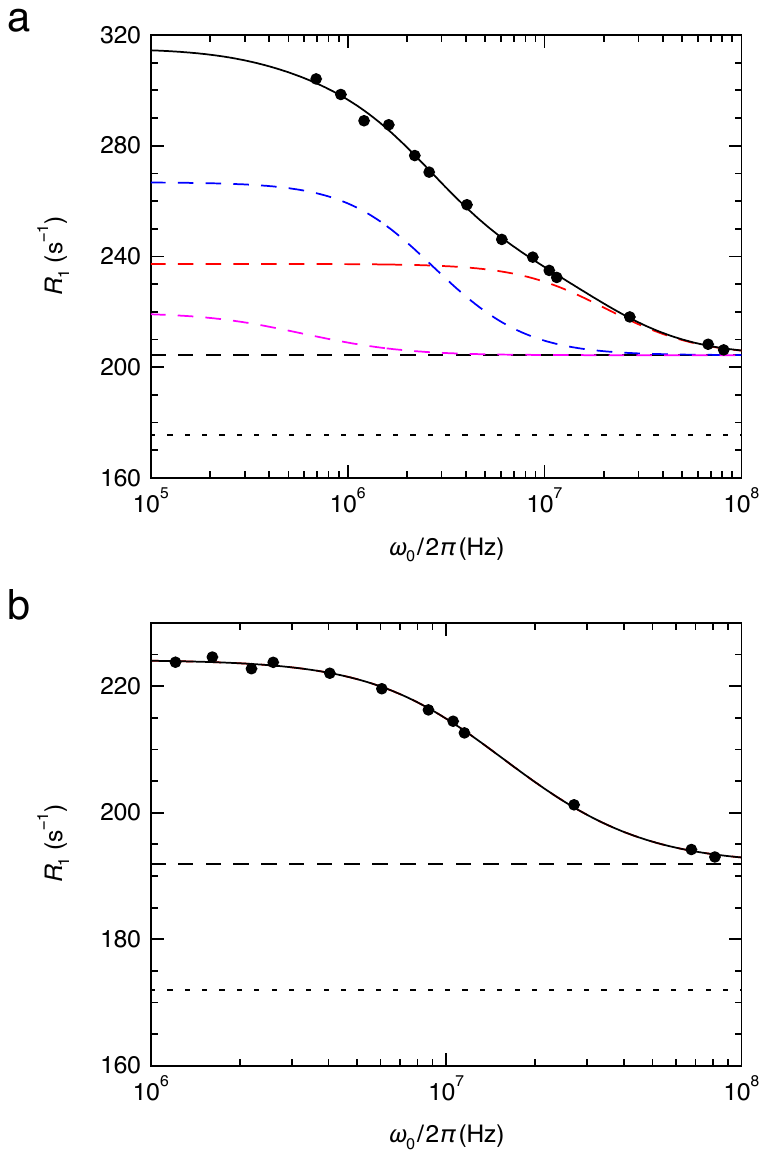}
  \caption{\label{fig4}(\textbf{a}) \osevt\ MRD profile, $R_1\nbr{\omega_0}$, 
           from MbCO gel at 25~$\degc$, pD 7.0, and $\nga$ = 29.9, scaled to 
           $N\rsub{W} = 3 \times 10^4$. 
           The error bars are up to twice the symbol size. 
           The solid curve is the fit that resulted in the parameter values in 
           Table~\ref{tab1}. 
           The dashed curves are dispersion components 1 (magenta), 2 (blue), 
           and~3 (red). 
           The bulk water (dotted line) and external hydration shell (dashed 
           line) contributions are shown in both panels. 
           (\textbf{b}) \osevt\ MRD profile, $R_1\nbr{\omega_0}$, from MbCO 
           solution at 25~$\degc$ and pD~7.0, scaled to 
           $N\rsub{W} = 3 \times 10^4$. 
           The error bars are comparable to the symbol size. 
           The solid curve is a single-component fit, with the parameters in 
           Table~\ref{tab1}.}
\end{figure}

Nevertheless, the \osevt\ MRD profile from \textit{immobilized} MbCO at pD~7.0 
exhibits a large dispersion in the examined frequency range (Fig.~\ref{fig4}a). 
In analyzing this MRD profile, we include the small contribution from 
component~1, taking $\tau\rsub{S,1} = 5.6$~\us\ and $S_1 = 0.79$ from the \htwo\ 
MRD fit (Table~\ref{tab1}) and using the estimate $N_1 = 5.2$ for the number of 
water molecules associated with this component (Sec.~\ref{sec3.2}). 
Even at the lowest frequency, this contribution hardly exceeds the measurement 
error in $R_1$ (Fig.~\ref{fig4}a). 
Beyond this minor contribution, two fast-exchange components are required to 
account for the \osevt\ MRD profile (Table~\ref{tab1}).

The dynamic perturbation factor $\xi\rsub{H}$ and the parameters 
$\tau\rsub{S,3}$ and $N_3 S\rsub{iso,3}^{\,2}$ of the fastest dispersion 
component do not differ significantly from the corresponding \htwo\ parameters 
(Table~\ref{tab1}), as expected if these contributions reflect weakly 
anisotropic water motions in the external hydration shell in the gel. 
The amplitude parameter, $N\rsub{iso,2}^{\,2} = 1.0 \pm 0.1$, of the slower 
\osevt\ component is similar to that of \htwo\ component~2 (Table~\ref{tab1}), 
but the factor four difference in the associated correlation times shows that 
these components have different physical origins. 
A water molecule with 120~ns MST would give rise to a very large \osevt\ 
dispersion, which is not observed, so we must conclude that this \htwo\ 
component is produced by LDs. 
Since 120~ns is too short to be an MST for a LD (all carboxyl groups are 
titrated at pD~7.0), we assign \htwo\ component~2 to internal motions of the His 
residues that contribute to \htwo\ component~1. 
Internal motions of protein side-chains on this time scale have previously been 
inferred from \htwo\ MRD studies of several proteins.\cite{Persson2008a,
VacaChavez2006b,Denisov1995b}

\osevt\ component~2, on the other hand, must reflect water motions. 
In principle, this could be local motions of the 5~\us\ internal waters. 
The only local water motion that might be as slow as 32~ns, however, is a 
180$\degree$ flip about the dipole axis, which, for symmetry reasons, cannot 
induce \osevt\ relaxation.\cite{Halle1999c} 
We therefore conclude that the dominant \osevt\ component~2 represents a single 
highly ordered water molecule with an MST of $32 \pm 3$~ns. 
This water molecule is unlikely to be deeply buried and it must exchange by a 
different mechanism than that used by the 5~\us\ waters. 
But why is this water molecule not evident as a 32~ns component in the \htwo\ 
profile? 
If it undergoes 180$\degree$ flips on the time scale 0.1 -- 10~\us, the \htwo\ 
parameter $S\rsub{iso}^{\,2}$ is reduced by a factor of three.\cite{Halle1999c} 
The contribution of this water molecule would then not be resolved because it 
would only add a small dispersion amplitude (0.3~$\pers$ in $R_1$) between the 
6~ns and 121~ns dispersion steps (Fig.~\ref{fig2}a).

As in the \htwo\ case, the \osevt\ MRD profile from \textit{free-tumbling} MbCO 
is well described by a single dispersion component (Fig.~\ref{fig4}b). 
The dynamic perturbation factor, $\xi\rsub{H} = 5.8 \pm 0.1$, is the same as 
obtained from the \htwo\ solution profile (Table~\ref{tab1}), as expected since 
water motions in the external hydration shell are nearly isotropic.\cite{
Halle1999c} 
The correlation time, $\tau\rsub{C} = 5.6 \pm 0.2$~ns, is shorter than the Mb 
tumbling time, $\tau\rsub{R} = 11.1$~ns, because the single \osevt\ solution MRD 
component is an unresolved superposition of two components, both with 
correlation times shorter than $\tau\rsub{R}$. 
The first component, due to water molecules with $\tau\rsub{S} = 5.6$~\us\ as 
deduced from the gel \htwo\ MRD profile, is in the intermediate-exchange regime 
for \osevt, with the effective correlation time shortened by a factor of 
$\sim 2$ according to Eq.~\eqref{eq3}. 
The contribution of this component to the total amplitude parameter, 
$N S\rsub{iso}^{\,2} = 3.0 \pm 0.1$ (Table~\ref{tab1}), is reduced by the same 
factor. 
The second component, due to the single water molecule with $\tau\rsub{S} = 
32$~ns as deduced from the gel \osevt\ MRD profile, is in the fast-exchange 
regime with effective correlation time 
$\nbr{1/\tau\rsub{R} + 1/\tau\rsub{S}}^{-1} = 8.3 \pm 0.2$~ns 
(Sec.~\ref{sec2.3}).

A quantitative analysis of the two contributions to 
$N S\rsub{iso}^{\,2} \tau\rsub{C}$, which is proportional to the magnitude of 
the dispersion, shows that the solution and gel \osevt\ MRD profiles are 
consistent, within the experimental error, with the gel \htwo\ MRD profile 
(Sec.~\ref{secS3}, Fig.~\ref{figS4}). 
Of particular interest is the number $N_1\rsup{W}$ of internal water molecules 
with an MST of 5.6~\us. 
Our analysis of the pD dependence of the gel \htwo\ MRD profile 
(Sec.~\ref{sec3.2}) indicates that $N_1\rsup{W} = 5.2$, but the \osevt\ MRD data 
suggest a somewhat smaller value (Sec.~\ref{secS3}, Fig.~\ref{figS4}). 
Averaging these results, we arrive at the conservative estimate $N_1\rsup{W} = 
4.5 \pm 1.0$. 

In summary, all the \htwo\ and \osevt\ MRD profiles measured on MbCO gels and 
solutions can be interpreted in a self-consistent way. 
The gel \htwo\ MRD profile and the solution \osevt\ MRD profile are both 
consistent with $N_1\rsup{W} = 4.5 \pm 1.0$ ordered water molecules with an MST 
of $5.6 \pm 0.5$~\us. 
The solution and gel \osevt\ MRD profiles indicate an additional component, not 
resolved in the \htwo\ profiles, comprising a single highly ordered water 
molecule with an MST of $32 \pm 3$~ns.

%
%
\subsection{\label{sec3.5}Internal hydration sites}

MRD experiments can provide the number of internal water molecules and their 
mean survival times, but the location of a hydration site can only be determined 
by MRD if the water molecule can be displaced by a ligand or a substituted 
side-chain.\cite{Halle1999c} 
Mb contains four cavities, denoted Xe1 -- Xe4 (Fig.~\ref{fig1}), that have been 
shown to bind xenon at elevated pressure.\cite{Tilton1984} 
To determine if the internal water molecules identified here occupy any of the 
Xe sites, we recorded the \htwo\ MRD profile from an MbCO gel that had been 
equilibrated with 8~bar Xe gas. 
We determined the Xe occupancies of these sites from \xe\ NMR spectra of the 
MbCO gel and of two reference samples (Fig.~\ref{figS5}), using also the 
reported\cite{Ewing1970} Xe binding constants for equine Mb at pH~7.0 and 
25~$\degc$. 
In this way, we obtained a Xe occupancy of $0.80 \pm 0.09$ for site Xe1 and a 
combined occupancy of $0.10 \pm 0.05$ for sites Xe2 -- Xe4 (Sec.~\ref{secS4}, 
Tables \ref{tabS3} and~\ref{tabS4}).

\begin{figure}[!t]
  \centering
  \includegraphics[viewport=194 342 401 500]{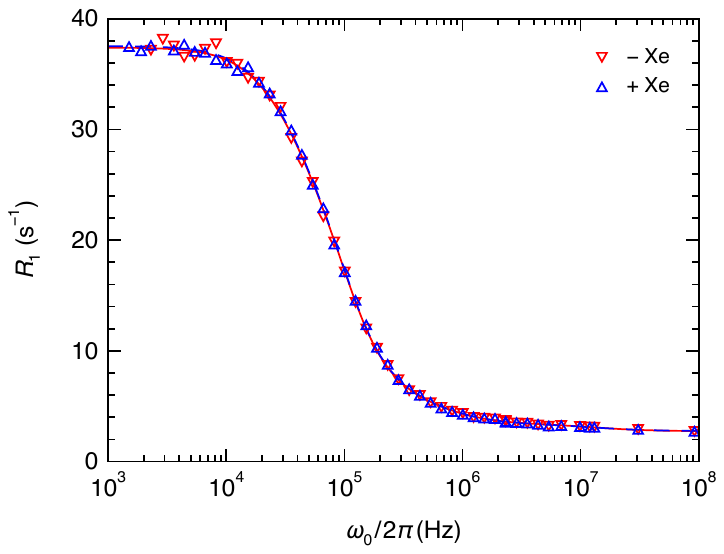}
  \caption{\label{fig5}\htwo\ MRD profile, $R_1\nbr{\omega_0}$, from immobilized 
           MbCO (25~$\degc$, pD~7.16, scaled to $N\rsub{W} = 3 \times 10^4$) in 
           the presence (blue symbols) or absence (red symbols) of 8~bar xenon. 
           The error bars do not exceed the symbol size. 
           The parameter values obtained from the fits (solid red and dashed 
           blue curves) are collected in Table~\ref{tabS5}.}
\end{figure}

As seen from Fig.~\ref{fig5} and Table~\ref{tabS5}, Xe binding at this level has no 
significant effect on the {\htwo} MRD profile. 
With 80~\% Xe occupancy in Xe1, we would have expected 
$100 \times 0.8/6.3 = 13$~\% reduction of $R_1$ at low frequency if Xe1 had been 
fully occupied by a long-lived water molecule.
We can therefore rule out internal water in this site. 
This is hardly surprising, since Xe1 has by far the highest Xe affinity\cite{
Ewing1970} among the four Xe sites and thus is expected to be the least polar 
site. 
For the other sites, the combined $10 \pm 5$~\% Xe occupancy is probably not 
sufficient to produce a detectable reduction of the MRD amplitude even if all 
these sites were fully occupied by water in the absence of Xe. 
If one Xe atom displaces one water molecule, MRD component~1 would be reduced by 
merely 1.5~\% (0.1/6.7), comparable to the measurement error.

All high-resolution crystal structures of equine and sperm whale Mb, in the CO, 
deoxy, or met forms,\cite{Maurus1997,Kachalova1999,Vojtechovsky1999,Chu2000,
Ostermann2002,Hersleth2007} identify two internal hydration sites in the Xe3 
cavity, with full (equine Mb) or nearly full (sperm whale Mb) occupancy (Figs.\ 
\ref{fig1} and~\ref{figS6}, Tables \ref{tabS6} and~\ref{tabS7}). 
The water molecule in subsite Xe3a has a small $B$ factor and is engaged in four 
H-bonds, donating to Ile-75 O and His-82 N$^{\updelta 1}$ and accepting from 
Gly-80 N and the adjacent water molecule. 
The water molecule in subsite Xe3b has a much larger $B$ factor and makes only 
one H-bond with the protein (Ala-134 O). 
In addition, all but one of the high-resolution crystal structures listed in 
Tables \ref{tabS6} and~\ref{tabS7} locate an internal water molecule in the 
so-called apical site (Figs.\ \ref{fig1} and~\ref{figS6}, Tables \ref{tabS6} 
and~\ref{tabS7}). 
The apical water makes three H-bonds with the protein. 
The crystal structures of equine Mb (but not sperm whale Mb) in the CO and deoxy 
(but not met) forms also locate a water molecule in the `bottom' site (Figs.\ 
\ref{fig1} and~\ref{figS6}, Tables \ref{tabS6} and~\ref{tabS7}). 
The bottom water also makes three H-bonds with the protein, but it is less 
deeply buried than the apical water.

It is tempting to identify these four crystallographic water molecules with the 
$4.5 \pm 1.0$ most long-lived (5~\us) water molecules indicated by the MRD data. 
The Xe difference MRD experiment does not rule out the presence of two water 
molecules in the Xe3 cavity because Xe appears to displace only one of these 
water molecules (at least for sperm whale Mb; see PDB entries 1J52 and 2W6Y) and 
then, as noted above, the effect would be too small to detect. 
Furthermore, the $10 \pm 5$~\% combined Xe occupancy is presumably associated 
mainly with the less polar Xe2 and Xe4 cavities. 
The 32~ns water inferred from the \osevt\ MRD data might be located in a deep 
surface pocket. 
Alternatively, it could refer to one of the two least deeply buried among the 
four internal hydration sites, the bottom site and subsite Xe3b. 
In that case, another internal hydration site, not identified in the crystal 
structures, must contribute to the 5~\us\ MRD component. 
For example, the phantom (Ph) cavity (Fig.~\ref{fig1}) is large enough to 
accommodate four (mutually H-bonded) water molecules, but this cavity is empty 
in all the crystal structures of equine Mb.

The two MD simulation studies that have focused on internal water in Mb both 
find multiple partially occupied hydration sites\cite{Scorciapino2010,
Lapelosa2013} most of which have not been identified in crystal structures. 
The first of these studies,\cite{Scorciapino2010} a conventional MD simulation 
of equine met-Mb at 300~K, found, in addition to the \ce{Fe^{3+}}-bound water in 
the DP, eight internal hydration sites with a combined occupancy of 2.75. 
The apical site (occupancy 0.8) was identified, but not the bottom site. 
Both the Xe3 and Xe4 cavities were assigned an occupancy of 0.3. 
Except for the iron-bound water, all internal water molecules identified from 
this 48~ns MD trajectory exchanged on a nanosecond time scale. 
Therefore, this MD study supports neither the crystallographically inferred 
locations and occupancies of internal hydration sites in Mb nor the MRD-inferred 
microsecond time scale of internal water exchange.

The second simulation study\cite{Lapelosa2013} examined sperm whale deoxy-Mb at 
310~K and used temperature-accelerated MD for enhanced sampling. 
Numerous internal hydration sites (but no occupancies) were reported, including 
the Xe2, Xe3, and Xe4 (but not Xe1) cavities. 
Based on computed minimum free energy pathways for water migration, it was 
suggested that all internal water molecules enter the protein via the His-64 
gate and the DP, from where water molecules can migrate further to the Xe 
cavities. 
With the aid of transition state theory, the time scale for water escape from 
the DP was estimated to 300~ns (at 310~K), an order of magnitude faster than the 
5~\us\ MST deduced from our MRD data.

In summary, crystallography\cite{Maurus1997,Kachalova1999,Vojtechovsky1999,
Chu2000,Ostermann2002,Hersleth2007} supports a scenario where most or all of the 
$\sim 4$ long-lived internal water molecules inferred by MRD reside in small 
polar cavities (Xe3, apical, and bottom sites). 
A larger number of partially occupied sites, as suggested by MD simulations, is 
not likely to account for the MRD data. 
X-ray crystallography may miss hydration sites with low occupancy and/or 
positional disorder, but water molecules in less polar sites (say, with less 
than three H-bonds) are expected to have a small order parameter and would then 
not contribute much to the MRD profile.

%
%
\subsection{\label{sec3.6}Mechanism of internal-water exchange}

The common MST, $\tau\rsub{S} = 5.6 \pm 0.5$~\us, of the $\sim 4$ water internal 
molecules inferred from our MRD data suggests that they exchange with external 
solvent via a common mechanism. 
Moreover, since the likely internal hydration sites are located far apart in the 
Mb molecule (the apical, bottom, and Xe3 sites are separated by 17, 27, and 
31~\AA), the common exchange mechanism must be global in spatial extent.

In an effort to gain more insight into the exchange mechanism, we studied the 
temperature dependence of the \htwo\ MRD profile from immobilized MbCO. 
Irreversible changes in the MRD profile, possibly caused by gel restructuring, 
were noted above 35~$\degc$. 
Therefore, the analysis was restricted to the rather narrow temperature range 
5 -- 25~$\degc$, where fits like the one in Fig.~\ref{fig2}a yielded a 
surprisingly small Arrhenius activation energy of $17.4 \pm 0.4$~kJ\;mol$^{-1}$ 
for the MST $\tau\rsub{S,1}$ at pD~7.0 (Fig.~\ref{figS7}). 
Such a small activation energy might suggest that water exchange is rate-limited 
by an entropic bottle-neck. 
The narrow temperature range and the strong covariance between $N_1$ and 
$\tau\rsub{S,1}$ (Sec.~\ref{sec3.2}), however, preclude an unambiguous 
interpretation of the temperature dependence. 
The MSTs used for the Arrhenius fit in Fig.~\ref{figS7} were obtained from 
constrained fits where $\tau\rsub{S,1}$ was the only adjustable parameter for 
component~1 even though the relative contributions of internal water and LDs to 
component~1 are expected to vary with temperature.

The $\sim 5$~\us\ MST deduced from our MRD analysis implies that internal-water 
exchange in Mb is a rare event that has not yet been captured by MD simulations, 
which so far do not extend beyond 100~ns for Mb. 
Neither the average protein structure seen in crystal structures nor the 
fluctuations accessed by sub-\us\ simulations tell us much about the transient 
structures involved in the water exchange event. 
An ultralong MD simulation\cite{Shaw2010} was recently used to demonstrate that 
the internal water molecules in the protein BPTI exchange by a transient 
aqueduct mechanism,\cite{Persson2013} where single-file water chains penetrate 
the protein through transiently formed tunnels or pores. 
It is conceivable that a similar mechanism operates in Mb, with the difference 
that the transient water-filled tunnels are partly built from preexisting 
cavities. 
Whereas a single water molecule is unlikely to migrate from a polar site into a 
largely nonpolar cavity network,\cite{Yu2010} an H-bonded water chain can move 
rapidly through a transiently formed tunnel.\cite{Rasaiah2008} 
The peripherally located internal hydration sites might then act both as portals 
for solvent penetration and as seeds for transient water chains.
 
MD simulations indicate that the ligands can migrate from the DP into the 
extensive network of permanent and transient cavities that permeates the Mb 
molecule.\cite{Cohen2006,Elber2008,Ruscio2008,Maragliano2010,Lin2011} 
The ligand migration pathways seen in these simulations include the apical and 
Xe3 hydration sites.\cite{Ruscio2008,Ceccarelli2008} 
A transient aqueduct mechanism might act to intermittently `flush' the cavity 
network, thereby removing internally trapped ligands. 
Experiments\cite{Olson1996,Scott2001,Nishihara2004,Goldbeck2006,Esquerra2008,
Cho2010} and simulations\cite{Ruscio2008} indicate that photo-dissociated CO can 
escape from the DP via the His-64 gate on a 100~ns time scale. 
The half-life of CO that has migrated from the DP to the Xe1 cavity was, 
however, reported to be $\sim 10$~\us\ (at 20~$\degc$).\cite{Srajer2001} 
Since this process occurs on the same time scale as internal-water exchange, it 
may, in fact, be concomitant with the proposed `flushing' of the cavity system 
by transient water chains.

%
%
\section{\label{sec4}Conclusions}

Based on \htwo\ and \osevt\ MRD measurements on gel and solution samples, we 
have arrived at the following answers to the first three of the five questions 
posed in Sec.~\ref{sec1}:
\begin{enumerate}[label=(\arabic*),series=conc]
  \item Under physiological conditions, equine MbCO contains $4.5 \pm 1.0$ 
        internal water molecules that exchange on a \us\ time scale. 
        The root-mean-square orientational order parameter of these water 
        molecules is 0.8. 
        Another ordered water molecule, possibly located in a deep surface 
        pocket, exchanges on a time scale of 30~ns.

  \item The likely locations of the $\sim 4$ long-lived water molecules are the 
        crystallographically identified Xe3, apical, and bottom hydration sites. 
        In agreement with crystallography and MD simulations, the MRD results 
        indicate that the Xe1 cavity is devoid of water.

  \item Despite being located as much as 30~\AA\ apart, the $\sim 4$ long-lived 
        internal water molecules have the same (or very similar) MST, 
        $\tau\rsub{S} = 5.6 \pm 0.5$~\us.
\end{enumerate}

\noindent In addition, the MRD results indicate that two or three of the 11 His 
residues of equine Mb undergo intramolecular hydrogen exchange on a \us\ time 
scale.

As regards the last two questions in the Introduction, we speculate that
\begin{enumerate}[label=(\arabic*),resume=conc]
  \item The internal water molecules exchange by a global mechanism, that may 
        involve penetration of the protein by transient H-bonded water chains, 
        entering the protein at the peripherally located permanent hydration 
        sites. 

  \item The proposed water exchange mechanism suggests a functional role, where 
        intermittent `flushing' of the cavity system removes trapped ligands. 
\end{enumerate}

%
%
\section*{Acknowledgments}

{\small
We thank Hanna Nilsson for help with sample purification. 
This work was financially supported by the Swedish Research Council. 
S.K.\ acknowledges postdoctoral fellowships from the Wenner-Gren Foundations and 
the Swedish Research Council. 
}

%
%

%
%
\onecolumn
\section*{Supporting information}

\setcounter{section}{0}
\setcounter{figure}{0}
\setcounter{table}{0}
\setcounter{equation}{0}

\renewcommand{\theHsection}{sup.\the\value{section}}
\renewcommand{\thesection}{S\arabic{section}}
\renewcommand{\thefigure}{S\arabic{figure}}
\renewcommand{\thetable}{S\arabic{table}}
\renewcommand{\theequation}{S\arabic{equation}}

\vspace*{\fill}
\begin{table}[!h]
  \centering
  \begin{threeparttable}
    \caption{\label{tabS1}Samples for MRD experiments.}
    \small
    \begin{tabular}{ccccclccc}
      \toprule
      \# & $C\rsub{Mb}$\tnote{a} & type\tnote{b} & $\nga$ & pD 
      & buffer\tnote{c} & nuclide & range (MHz) & $n$\tnote{d} \\
      \midrule
      1           & 1.98 & S &      & 5.06 &              & \htwo  
      & 0.00152 -- 92.1 & 34 \\
      2           & 2.43 & S &      & 6.00 &              & \htwo  
      & 0.00152 -- 92.1 & 34 \\
      3           & 2.35 & S &      & 7.00 &              & \htwo  
      & 0.00152 -- 92.1 & 34 \\
      4           & 2.38 & S &      & 8.03 &              & \htwo  
      & 0.00152 -- 92.1 & 34 \\
      5           & 2.28 & S &      & 9.18 &              & \htwo  
      & 0.00152 -- 92.1 & 34 \\
      6           & 2.22 & S &      & 9.97 &              & \htwo  
      & 0.00152 -- 92.1 & 34 \\
      \midrule
      7           & 1.02 & G & 110  & 5.95 & 5~\mM\ NaPi   & \htwo  
      & 0.00152 -- 5.37 & 40 \\
      8           & 1.63 & G & 62.1 & 5.66 & 5~\mM\ NaPi   & \htwo  
      & 0.00152 -- 5.37 & 40 \\
      9           & 1.67 & G & 31.1 & 5.73 & 5~\mM\ NaPi   & \htwo  
      & 0.00152 -- 5.37 & 40 \\
      10          & 5.14 & G & 30.5 & 5.79 & 5~\mM\ NaPi   & \htwo  
      & 0.00152 -- 5.37 & 40 \\
      11          & 1.35 & G & 30.5 & 6.03 & 5~\mM\ NaPi   & \htwo  
      & 0.00152 -- 5.37 & 40 \\
      12          & 2.65 & G & 30.4 & 5.90 & 5~\mM\ NaPi   & \htwo  
      & 0.00152 -- 5.37 & 40 \\
      13          & 1.62 & G & 30.3 & 6.41 & 50~\mM\ NaPi  & \htwo  
      & 0.00152 -- 5.37 & 40 \\
      14          & 1.61 & G & 30.5 & 6.23 & 5~\mM\ NaPi   & \htwo  
      & 0.00152 -- 5.37 & 40 \\
      15          & 1.61 & G & 30.4 & 6.39 & 50~\mM\ NaPi  & \htwo  
      & 0.00152 -- 5.37 & 40 \\
      16          & 1.60 & G & 30.6 & 6.70 & 50~\mM\ NaPi  & \htwo  
      & 0.00152 -- 5.37 & 40 \\
      17          & 1.60 & G & 30.6 & 6.92 & 50~\mM\ NaPi  & \htwo  
      & 0.00152 -- 5.37 & 40 \\
      18          & 1.60 & G & 30.6 & 6.87 & 50~\mM\ NaPi  & \htwo  
      & 0.00152 -- 5.37 & 40 \\
      19          & 1.61 & G & 30.4 & 6.09 & 50~\mM\ NaPi  & \htwo  
      & 0.00152 -- 5.37 & 40 \\
      20          & 1.61 & G & 30.4 & 6.75 & 50~\mM\ MES   & \htwo  
      & 0.00152 -- 92.1 & 45 \\
      21          & 1.61 & G & 30.4 & 6.13 & 50~\mM\ NaPi  & \htwo  
      & 0.00152 -- 5.37 & 40 \\
      22          & 1.65 & G & 29.7 & 6.12 & 50~\mM\ MES   & \htwo  
      & 0.00152 -- 5.37 & 40 \\
      23          & 1.64 & G & 29.9 & 5.99 & 50~\mM\ MES   & \htwo  
      & 0.00152 -- 5.37 & 40 \\
      24          & 1.65 & G & 29.8 & 6.35 & 50~\mM\ MES   & \htwo  
      & 0.00152 -- 5.37 & 40 \\
      25\tnote{e} & 1.65 & G & 29.8 & 6.64 & 50~\mM\ MES   & \htwo  
      & 0.00152 -- 5.37 & 40 \\
      26          & 1.61 & G & 30.4 & 6.75 & 50~\mM\ MES   & \htwo  
      & 0.00152 -- 92.1 & 45 \\
      27          & 1.62 & G & 29.9 & 6.98 & 50~\mM\ PIPES & \htwo  
      & 0.00152 -- 5.37 & 40 \\
      28\tnote{e} & 1.62 & G & 29.9 & 6.95 & 50~\mM\ PIPES & \htwo  
      & 0.00152 -- 5.37 & 40 \\
      29          & 1.62 & G & 29.9 & 7.00 & 50~\mM\ PIPES & \htwo  
      & 0.00152 -- 92.1 & 45 \\
      30          & 1.62 & G & 29.9 & 7.15 & 50~\mM\ PIPES & \htwo  
      & 0.00152 -- 92.1 & 46 \\
      31\tnote{f} & 1.62 & G & 29.9 & 7.17 & 50~\mM\ PIPES & \htwo  
      & 0.00152 -- 92.1 & 46 \\
      \midrule
      32          & 2.57 & S &      & 7.03 &               & \osevt 
      & 1.21 -- 81.3    & 12 \\
      \midrule
      33          & 1.62 & G & 29.9 & 7.02 & 50~\mM\ PIPES & \osevt 
      & 0.692 -- 81.3   & 14 \\
      \bottomrule
    \end{tabular}
    \begin{tablenotes}[flushleft,para]
      \footnotesize
      \item[a] MbCO concentration in \mM.
      \item[b] Solution (S) or gel (G).
      \item[c] NaPi $=$ sodium phosphate.
      \item[d] Number of data points.
      \item[e] Measured at 5, 15, and~25~$\degc$.
      \item[f] In presence of 8~bar Xe.
    \end{tablenotes}
  \normalsize
  \end{threeparttable}
\end{table}
\vspace*{\fill}

\newpage

\begin{table}[!h]
  \centering
  \begin{threeparttable}
    \caption{\label{tabS2}$\pka^\ast$ values for titrating His residues in horse 
           Mb.\tnote{a}}
    \small
    \begin{tabular}{cccc}
      \toprule
      \multirow{3}{*}{residue} & \multicolumn{3}{c}{$\pka^\ast$} \\
      \cmidrule(lr){2-4}
      & 299.5~K, 0.2~\M\ NaCl\cite{Bhattacharya1997} 
      & 298~K, 20~\mM\ NaCl\cite{Kao2000} 
      & 298~K, 0.2~\M\ NaCl\cite{Kao2000} \\
      \midrule
      36  & $7.75 \pm 0.05$ & $7.67 \pm 0.01$ & $7.80 \pm 0.02$ \\
      48  & $5.50 \pm 0.07$ & $5.42 \pm 0.02$ & $5.62 \pm 0.01$ \\
      81  & $6.86 \pm 0.05$ & $6.65 \pm 0.01$ & $6.94 \pm 0.01$ \\
      113 & $5.76 \pm 0.07$ & $5.51 \pm 0.02$ & $5.87 \pm 0.02$ \\
      116 & $6.72 \pm 0.05$ & $6.66 \pm 0.02$ & $6.79 \pm 0.01$ \\
      119 & $6.51 \pm 0.05$ & $6.38 \pm 0.02$ & $6.56 \pm 0.01$ \\
      \bottomrule
    \end{tabular}
    \begin{tablenotes}[flushleft]
      \footnotesize
      \item[a] The number of labile His deuterons was calculated as 
               $N\rsub{LD} = 2 \sum_i \sbr{1 + 10^{\mathrm{pH}^\ast - 
               \mathrm{p}K^\ast_{\mathrm{a},i}}}^{-1}$, where pH$^\ast$ and 
               $\mathrm{p}K^\ast_{\mathrm{a},i}$ are both measured in 
               \ce{D2O} and neither are corrected for the H/D isotope effect 
               (which largely cancels out in the difference). 
               The set of six $\pka^\ast$ values used for the calculation 
               in Sec.~\ref{secS1} were measured\cite{Kao2000} at 298~K in 
               20~\mM\ NaCl. 
               In addition, we used $\pka^\ast = 5.91$ for His-97, 
               determined for sperm whale MbCO.\cite{Muller1999} 
               The remaining four His residues are all in the basic from in 
               the neutral pD range of interest here.\cite{Muller1999,
               Rabenstein2001}
     \end{tablenotes}
   \normalsize
   \end{threeparttable}
\end{table}

%
%
\section{\label{secS1}Histidine labile-deuteron exchange kinetics}

We estimate the mean survival time, $\tau\rsub{LD}$, of the acidic imidazolium 
deuterons from
\begin{equation}
  \tau\rsub{LD} = \frac{1}{k\rsub{w}\sbr*{\text{\ce{D2O}}} + 
  k\rsub{b}\sbr*{\text{\ce{OD-}}} + k\rsub{c}\sbr*{\ce{Buf-}}}\ ,
  \label{eqS1}
\end{equation}
where the three rate constants refer to the water, hydroxide, and buffer 
catalyzed exchange processes:
\begin{center}
  \begin{tabular}{lcl}
    \ce{$\geqslant$ND+ + D2O} & \ce{->[k\rsub{w}]} &\ce{$\geqslant$N{\text{:}} 
    + D3O+}\\
    \ce{$\geqslant$ND+ + OD-} & \ce{->[k\rsub{b}]} &\ce{$\geqslant$N{\text{:}} 
    + D2O} \\
    \ce{$\geqslant$ND+ + Buf-}& \ce{->[k\rsub{c}]} & \ce{$\geqslant$N{\text{:}} 
    + BufD} .\\
  \end{tabular}
\end{center}

Assuming that association and dissociation are diffusion-controlled with the 
rate constant $k\rsub{D}$, the three LD exchange rate constants can be expressed 
as\cite{Eigen1964}
\begin{equation}
  k = \frac{k\rsub{D}}{1+10^{\pka\nbr*{\mathrm{His}}-\pka\nbr*{\mathrm{BD}}}}\ ,
  \label{eqS2}
\end{equation}
where $\pka\nbr{\mathrm{His}}$ and $\pka\nbr{\mathrm{BD}}$ are the $\pka$ of the 
acidic form of the His side-chain and of the conjugate acid BD of the base that 
accepts the deuteron from His, that is \ce{D3O+}, \ce{D2O}, or BufD, 
respectively. 
For the following estimates, we ignore the H/D isotope effect, which largely 
cancels out in the $\pka$ difference in Eq.~\eqref{eqS2}.

Taking $k\rsub{D} = 10^{10}$~\M$^{-1}\;\pers$ (probably an overestimate), 
$\pka\nbr{\mathrm{His}} = 6.38$ (the average of the values in the second 
$\pka^\ast$ column in Table~\ref{tabS2}), $\pka\nbr{\text{\ce{H3O+}}} = -1.74$, 
and $sbr{\text{\ce{D2O}}} = 55$~\M, we obtain 
$k\rsub{w}^\prime = k\rsub{w} \sbr{\text{\ce{D2O}}} = 4.2 \times 10^3~\pers 
\approx \nbr{240~\text{\us}}^{-1}$. 
This estimate is close to the value, $k\rsub{w}^\prime = 2.4 \times 10^3~\pers$, 
measured by NMR for imidazole in \ce{H2O} at 25~$\degc$.\cite{Ralph1969}  
For the hydroxide-catalyzed process, with $\pka\nbr{\text{\ce{H2O}}} = 15.74$, 
we obtain the estimate $k\rsub{b} = k\rsub{D} = 10^{10}~\text{\M}^{-1}\;\pers$. 
With $\pka\nbr{\text{\ce{D2O}}} = 14.95$ at 25~$\degc$, we then obtain, at 
pD~7.0, $k\rsub{b} \sbr{\text{\ce{OD-}}} = 10^{10} \times 10^{7.0-14.95}~\pers 
\approx 1.1 \times 10^2~\pers = \nbr{9~\mathrm{ms}}^{-1}$.

To make a significant contribution to the dominant component of the \htwo\ MRD 
profile (Fig.~\ref{fig2}a), with a correlation time of $\sim 5$~\us, the mean 
survival time $\tau\rsub{LD}$ of the labile His deuterons must also be 
$\sim 5$~\us. 
Our rough but conservative estimates show that the water and hydroxide catalyzed 
processes are too slow to contribute significantly. 
The buffer catalyzed process can also be ruled out, since a tenfold increase of 
the phosphate buffer concentration (from 5 to 50~\mM) has no significant effect 
on $R_1\nbr{0}$ (Fig.~\ref{fig2}b). 
Furthermore, at 50~\mM\ buffer, no variation in $R_1\nbr{0}$ was observed 
between the phosphate ($\pka = 7.20$), PIPES ($\pka = 6.76$), and MES 
($\pka = 6.15$) buffers.

\begin{figure}[!h]
  \centering
  \includegraphics[viewport=104 368 491 474]{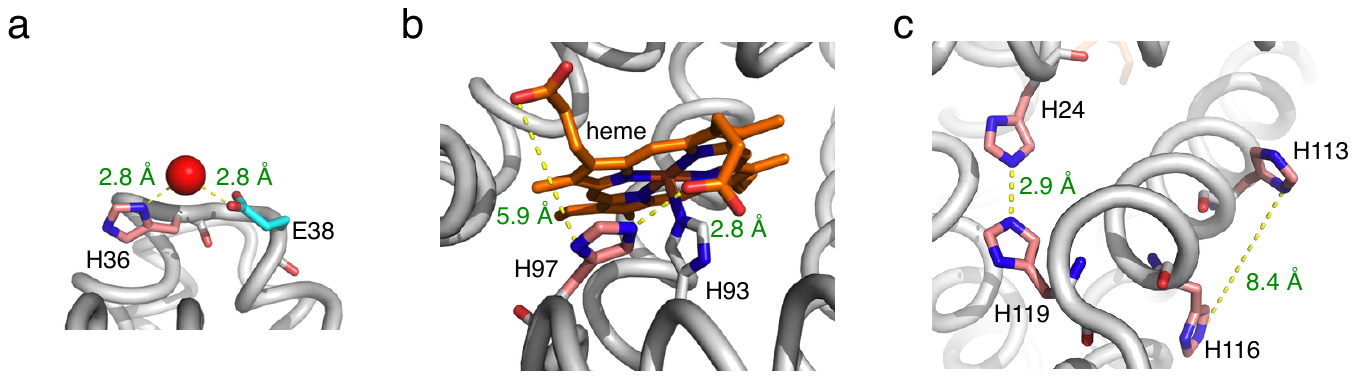}
  \caption{\label{figS1}Possible internal catalysis of labile-deuteron exchange 
           for (\textbf{a}) His-36, (\textbf{b}) His-97, and (\textbf{c}) 
           His-24/His-119 and His-113/His-116. 
           A red sphere represents a crystallographically identified water 
           molecule.\cite{Chu2000}}
\end{figure}

\begin{figure}[!h]
  \centering
  \includegraphics[viewport=180 342 415 500]{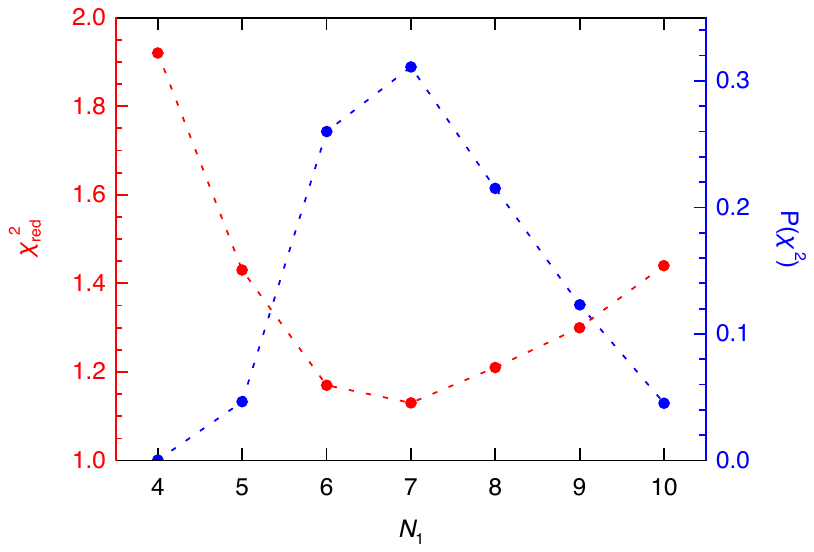}
  \caption{\label{figS2}Reduced $\chi^2$ and $P\nbr{\chi^2}$ for constrained 
           fits to \htwo\ MRD profile from immobilized MbCO at pD~7.0 with $N_1$ 
           for component~1 fixed in the fit.}
\end{figure}

\begin{figure}[!h]
  \centering
  \includegraphics[viewport=201 343 397 499]{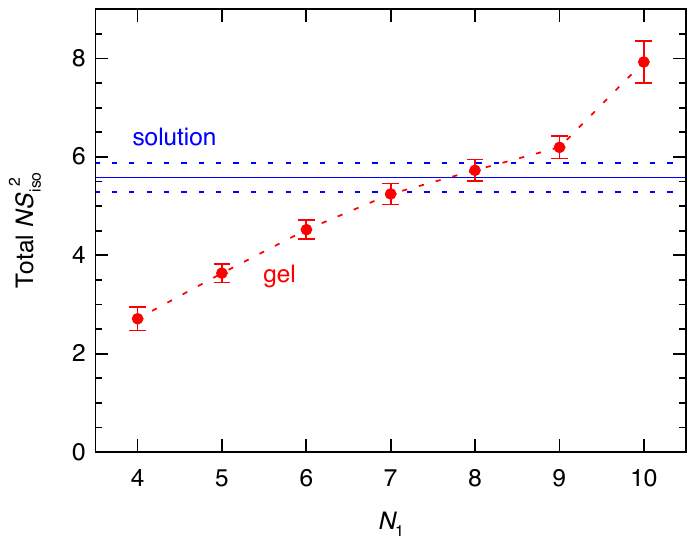}
  \caption{\label{figS3}Amplitude parameter $N S\rsub{iso}^2$ from 
           single-component fit to the solution \htwo\ MRD profile at pD~7.0 
           (blue line) and computed from the parameters of components 1 and~2 
           in constrained fits (fixed $N_1$) to the gel \htwo\ MRD profile at 
           pD~7.0 (red circles).}
\end{figure}

A remaining possibility is an internal catalysis such as 
\ce{$\geqslant$ND^{+}$\cdots$O_{w}D$\cdots$^{-}OOC}, where a water molecule 
connects the acidic His side-chain with a nearby proton acceptor, such as a 
carboxylate group or a basic His side-chain. 
The crystal structure\cite{Chu2000} 1DWR of equine MbCO suggests several such 
possibilities (Fig.~\ref{figS1}). 
For example, His-36 has an upshifted $\pka$ (Table~\ref{tabS2}),\cite{
Bhattacharya1997,Kao2000} presumably due to the proximity of Glu-38, the 
carboxylate group of which is bridged to His-36 via a H-bonding water molecule 
(Fig.~\ref{figS1}a).

%
%
\section{\label{secS2}pD-dependent effective correlation time}

The overall correlation time, $\tau\rsub{C}$, deduced from one-component fits to 
solution \htwo\ MRD profiles (Fig.~\ref{fig3}), decreases with pD even though 
$\tau\rsub{C}\rsup{eff}$ for the LDs increases with pD according to 
Eq.~\eqref{eq3}. 
This happens because LDs make a larger relative contribution to the MRD profile 
at higher pD. 
To understand this qualitatively, consider a single class of LDs with 
pD-dependent mean survival time $\tau\rsub{S}\nbr{\mathrm{pD}}$. 
In view of Eq.~\eqref{eq3}, we then have approximately
\begin{equation}
  \frac{\tau\rsub{C}}{\tau\rsub{R}} = \frac{N\rsup{W} + N\rsup{LD}/A}
  {N\rsup{W} + N\rsup{LD}/\sqrt{A}}\ ,
  \label{eqS3}
\end{equation}
with 
\begin{equation}
  A = 1 + \omega\rsub{Q}^{\,2} S\rsub{iso}^{\,2} \tau\rsub{R} 
  \tau\rsub{S}\nbr*{\mathrm{pD}}\ .
  \label{eqS4}
\end{equation}
This heuristic expression predicts that $\tau\rsub{C} = \tau\rsub{R}$ at low pD, 
where $\tau\rsub{S}$ is so long that the LD contribution is negligible compared 
to the internal-water contribution. 
With increasing pD, $\tau\rsub{S}$ is shortened by base-catalyzed LD exchange. 
As a result, $\tau\rsub{C}$ becomes progressively shorter than $\tau\rsub{R}$, 
as predicted by Eq.~\eqref{eqS3}. 
At still higher pD, $\tau\rsub{C}$ exhibits a minimum and finally, when 
$\tau\rsub{S}$ is so short that $A = 1$, $\tau\rsub{C}$ again approaches 
$\tau\rsub{R}$. 
In practice, the minimum is not observed (at least not in our pD range) because 
new classes of LDs with smaller exchange rate constants gradually come into play 
as pD is increased.

\newpage

%
%
\section{\label{secS3}Mutual consistency of \htwo\ and \osevt\ MRD results}

Here, we show that the parameters of the single component in the solution 
\osevt\ MRD profile is consistent with the results deduced from the gel \htwo\ 
and \osevt\ MRD profiles. 
To a good approximation, the product of the amplitude parameter and correlation 
time for the single \osevt\ solution MRD component can be expressed in terms of 
the parameters of the two components (identified in the gel MRD profiles) that 
it comprises:
\begin{equation}
    N S\rsub{iso}^{\,2} \tau\rsub{C} = \frac{N_1\rsup{W} S\rsub{iso,1}^{\,2} 
    \tau\rsub{R}}{1 + \omega\rsub{Q}^{\,2} S\rsub{iso,1}^{\,2} \tau\rsub{R} 
    \tau\rsub{S,1}} + N_2 S\rsub{iso,2}^{\,2} \sbr*{\frac{1}{\tau\rsub{R}} 
    + \frac{1}{\tau\rsub{S,2}}}^{-1}\ .
    \label{eqS5}
\end{equation}
From the fit results in Table~\ref{tab1}, we know that 
$N S\rsub{iso}^{\,2} \tau\rsub{C} = 16.8 \pm 0.8$~ns and that the second term 
on the right-hand side of Eq.~\eqref{eqS5} equals $8.3 \pm 0.8$~ns. 
The two components thus make comparable contributions to the solution \osevt\ 
MRD profile. 
In the first term on the right-hand side of Eq.~\eqref{eqS5}, we know that 
$\omega\rsub{Q} = 7.61\times 10^6$~rad\;$\pers$, $\tau\rsub{R} = 11.1$~ns, and 
$\tau\rsub{S,1} = 5.6$~\us\ (Table~\ref{tab1}). 
In Eq.~\eqref{eqS5}, $S\rsub{iso,1}$ refers to the \osevt\ order parameter, 
which may differ somewhat from the \htwo\ order parameter, $S\rsub{iso,1} = 
S_1\nbr{1+\eta_1^{\,2}/3}^{1/2} = 0.80 \pm 0.02$ (Table~\ref{tab1}). 
The \htwo\ order parameter pertains to internal water molecules as well as to 
rapidly exchanging LDs in a few His residues (Sec.~\ref{sec3.2}).

\begin{figure}[!h]
  \centering
  \includegraphics[viewport=198 343 399 499]{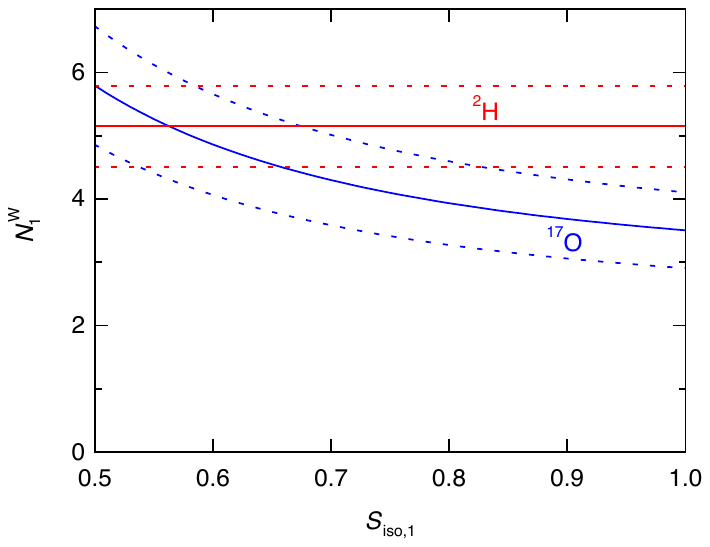}
  \caption{\label{figS4}The number $N_1\rsup{W}$ of internal water molecules 
           with MST 5.6~\us, deduced with the aid of Eq.~\eqref{eqS5} from the 
           solution and gel \osevt\ MRD profiles, as a function of the \osevt\ 
           order parameter $S\rsub{iso,1}$ (blue curve). 
           Also shown is the $N_1\rsup{W}$ value obtained from the pD-dependent 
           gel \htwo\ MRD profile (red line). 
           The dotted curves or lines indicate the propagated measurement 
           error.}
\end{figure}

From our analysis of the pD dependence of the gel \htwo\ MRD profiles, we 
estimated that component~1 comprises $N_1\rsup{W} = 5.2 \pm 0.6$ internal water 
molecules (Sec.~\ref{sec3.2}). 
This result might be affected by modest systematic errors due to the strong 
covariance of $N_1$ and $\tau\rsub{S,1}$ in the gel \htwo\ MRD fit and to 
inaccuracies in our analysis of the His LD contribution (Sec.~\ref{sec3.2}). 
If our analysis of the gel \htwo\ and \osevt\ MRD profiles is quantitatively 
consistent, Eq.~\eqref{eqS5} should be satisfied with $N_1\rsup{W} = 5.2$ and a 
reasonable \osevt\ order parameter $S\rsub{iso,1}$. 
Figure~\ref{figS4} shows that Eq.~\eqref{eqS5} is satisfied for \osevt\ 
$S\rsub{iso,1}$ values from below 0.5 up to 0.83 when the propagated measurement 
errors are taken into account. 
Nonetheless, over most of this $S\rsub{iso,1}$ range, Eq.~\eqref{eqS5} predicts 
a smaller $N_1\rsup{W}$ value than the one deduced from the pD dependence of the 
gel \htwo\ MRD profiles. 
The \osevt\ order parameter is not likely to be substantially smaller than the 
\htwo\ order parameter. 
Assuming that $S\rsub{iso,1} = 0.80$ also for \osevt, we obtain 
$N_1\rsup{W} = 3.9 \pm 0.6$ from Eq.~\eqref{eqS5} (Fig.~\ref{figS4}). 
Taking all available MRD data into account by averaging these two $N_1\rsup{W}$ 
values, we arrive at the more conservative estimate $N_1\rsup{W} = 4.5 \pm 1.0$.

%
%
\section{\label{secS4}Determination of xenon occupancies}

The Xe binding constants for equine Mb in the ferrous state at pH~7.0 and 
25~$\degc$ have been reported as $K_1 = 94$~\M$^{-1}$ for the Xe1 site and $K_2 
= 2.6$~\M$^{-1}$ for the weaker sites Xe2 -- Xe4 (treated as one 
site).\cite{Ewing1970}
We compute the Xe occupancy as
\begin{equation}
  \theta_n = \frac{K_n \sbr*{\mathrm{Xe}}}{1 + K_n \sbr*{\mathrm{Xe}}}\ ,
  \label{eqS6}
\end{equation}
where $n =$ 1 or 2 and [Xe] is the free Xe concentration. We used \xe\ NMR to 
determine the total Xe concentration 
$C\rsub{Xe} = \sbr{\mathrm{Xe}} + \theta C\rsub{Mb}$, where 
$\theta = \theta_1 + \theta_2$ and $C\rsub{Mb} = 1.62$~\mM\ is the total MbCO 
concentration in the gel sample.

\begin{figure}[!h]
  \centering
  \includegraphics[viewport=0 0 214 162]{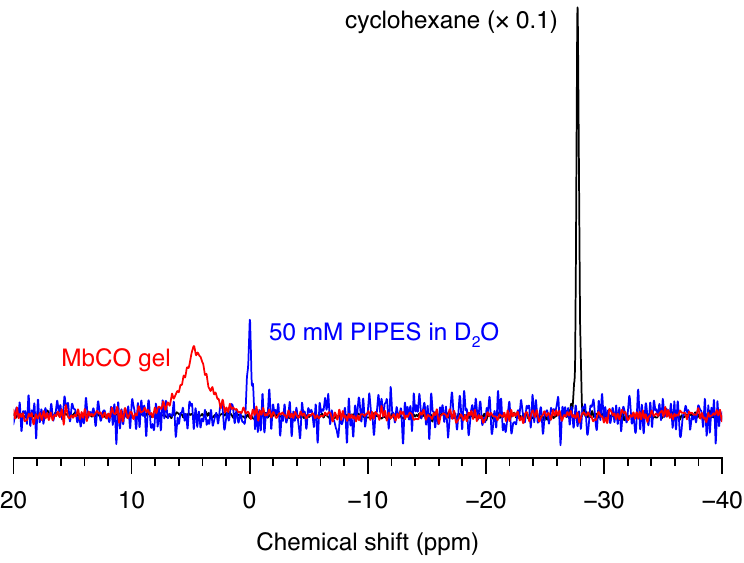}
  \caption{\label{figS5}Natural abundance \xe\ spectra recorded at 14.4~T and 
           25.0~$\degc$ from MbCO gel (8~bar Xe) and from the two reference 
           samples, PIPES buffer and cyclohexane (1~atm Xe). 
           The spectra are scaled to the same $N\rsub{S} = 256$ and, in addition, 
           the spectrum from cyclohexane was multiplied by 0.1. 
           The \xe\ chemical shift scale is referenced to the PIPES buffer 
           sample.}
\end{figure}

\begin{figure}[!h]
  \centering
  \includegraphics[viewport=93 330 502 512]{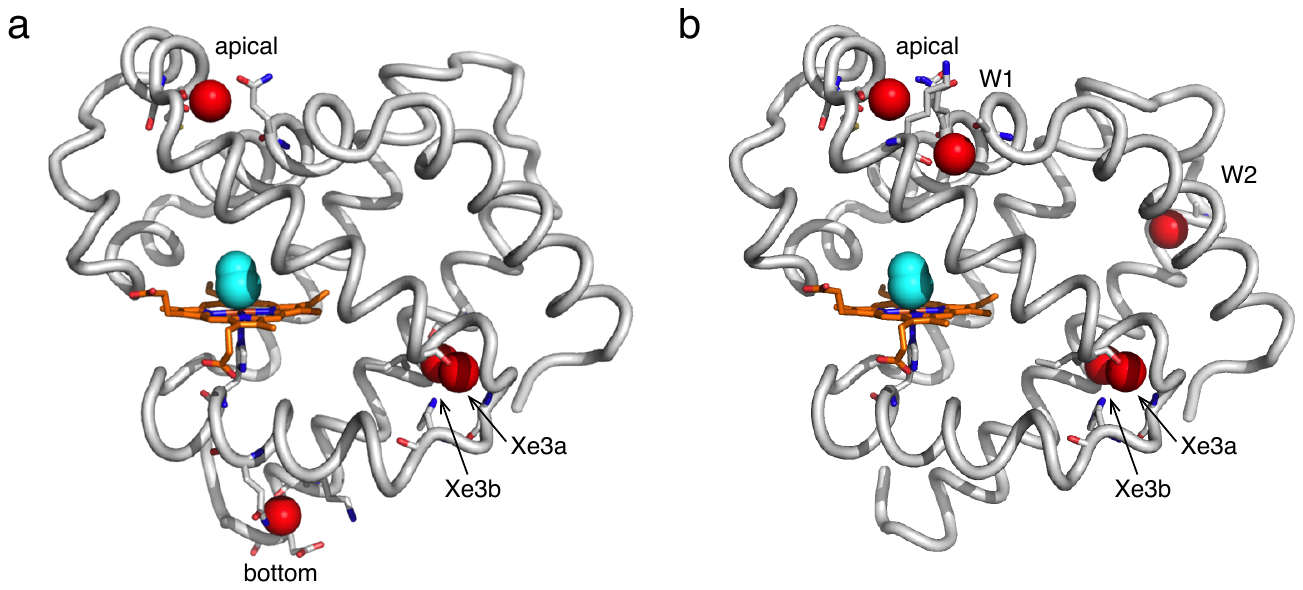}
  \caption{\label{figS6}Internal water molecules (red spheres) in crystal 
           structures of (\textbf{a}) equine MbCO (PDB:~1DWR\cite{Chu2000}) and 
           (\textbf{b}) sperm whale MbCO (PDB:~1A6G\cite{Vojtechovsky1999}) at 
           cryogenic temperature.}
\end{figure}

\begin{figure}[!h]
  \centering
  \includegraphics[viewport=199 343 397 499]{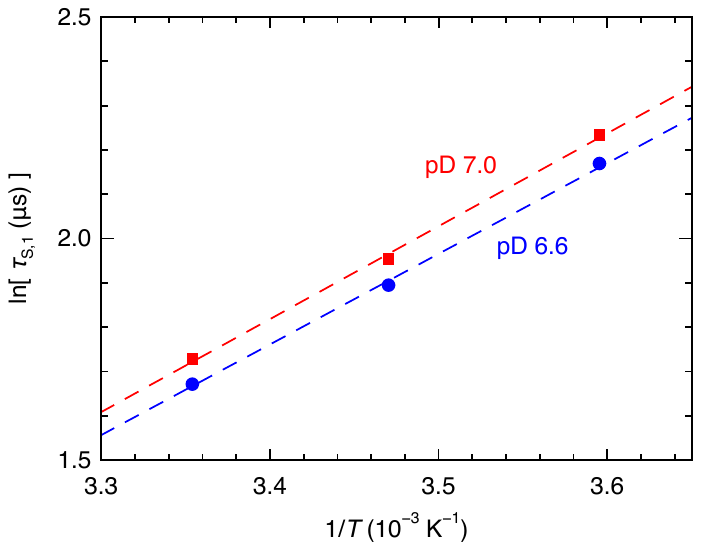}
  \caption{\label{figS7}Arrhenius plot of the MST $\tau\rsub{S,1}$ for 
           component~1 in the \htwo\ MRD profile from immobilized MbCO at the 
           indicated pD values. 
           The slope yields $E\rsub{A} = 17.4 \pm 0.4$~kJ\;mol$^{-1}$ at pD~7.0 
           and $17.0 \pm 0.3$~kJ\;mol$^{-1}$ at pD 6.6.}
\end{figure}

To determine $C\rsub{Xe}$ in the MbCO gel sample, we recorded \xe\ NMR spectra 
(Fig.~\ref{figS5}) from the gel equilibrated with 8~bar Xe (Sec.~\ref{sec2.1}) 
and, for calibration purposes, from two reference samples (50~\mM\ PIPES buffer 
in \ce{D2O} at pD~7.4 and cyclohexane) equilibrated with 1~atm Xe. 
The integrated intensity $I\rsub{Xe}$ of the Xe peak is proportional to the 
total Xe concentration $C\rsub{Xe}$ in the sample, but it also depends on the 
sample volume $V$ (in the active region of the RF coil), the number $N\rsub{S}$ 
of accumulated scans (transients), the \xe\ longitudinal relaxation rate 
$R_1\rsup{Xe}$, and the recycle delay $\tau\rsub{RD}$ according to 
\begin{equation}
  I\rsub{Xe} \propto C\rsub{Xe} V N\rsub{S} \sbr*{1 - 
  \exp\nbr*{- R_1\rsup{Xe} \tau\rsub{RD}}}\ .
  \label{eqS7}
\end{equation}
The values of these parameters, along with the reported Xe concentration in the 
reference solvents (at 25~$\degc$ and 1~atm Xe) are collected in 
Table~\ref{tabS3}. 
The resulting total Xe concentrations $C\rsub{Xe}$ are given in 
Table~\ref{tabS4}, with errors propagated from the quoted intensity errors and 
an estimated 50~\% uncertainty in $R_1\rsup{Xe}$ for the MbCO gel sample. 
The Xe occupancies $\theta_1$ and $\theta_2$ were then computed iteratively with 
the aid of Eq.~\eqref{eqS6}. 
In the main text, we use occupancies that are the averages of the values 
obtained with the two reference samples.

\vspace*{\fill}

\vspace*{\fill}

\begin{table}[!h]
  \centering
  \begin{threeparttable}
    \caption{\label{tabS3}Parameters used for determining Xe concentrations.}
    \small
    \begin{tabular}{lcccccc}
      \toprule
      sample & $I$ (a.u.) & $C\rsub{Xe}$ (\mM) & $V$ (\ul) & $N\rsub{S}$ & 
      $\tau\rsub{RD}$ (s) & $R_1\rsup{Xe}$ ($\pers$) \\
      \midrule
      MbCO gel      & $105.5 \pm 1.1$   & --           & 530 & 2560 & 
      20.0264  & 0.033\tnote{a} \\
      50 \mM\ PIPES & $1.678 \pm 0.185$ & 4.5\tnote{b} & 720 & 256  & 
      300.0264 & $1.98\times10^{-3}$\tnote{\,c} \\
      cyclohexane   & $33.77 \pm 0.083$ & 192\tnote{d} & 710 & 128  & 
      60.0264  & 0.017\tnote{e} \\
      \bottomrule
    \end{tabular}
    \begin{tablenotes}[flushleft,para]
      \footnotesize
      \item[a] Estimated from \xe\ spectra with different $\tau\rsub{RD}$.
      \item[b] Ref.~\citenum{Wilhelm1977}.
      \item[c] Ref.~\citenum{Rubin2000}.
      \item[d] Ref.~\citenum{Pollack1989}.
      \item[e] Ref.~\citenum{Oikarinen1995}.
    \end{tablenotes}
  \normalsize
  \end{threeparttable}
\end{table}

\vspace*{\fill}

\begin{table}[!h]
  \centering
  \begin{threeparttable}
    \caption{\label{tabS4}Xenon concentration and site occupancies in MbCO gel 
             sample.}
    \small
    \begin{tabular}{lccc}
      \toprule
      \multirow{3}{*}{reference} & \multirow{3}{*}{$C\rsub{Xe}$ (\mM)} 
      & \multicolumn{2}{c}{occupancy, $\theta_n$} \\
      \cmidrule(lr){3-4}
      & & Xe1 & Xe2 -- Xe4 \\
      \midrule
      50 \mM\ PIPES & $34.2 \pm 13.2$ & $0.76 \pm 0.07$ & $0.08 \pm 0.03$ \\
      cyclohexane   & $51.6 \pm 18.8$ & $0.83 \pm 0.05$ & $0.12 \pm 0.04$ \\
      average       &                 & $0.80 \pm 0.09$ & $0.10 \pm 0.05$ \\
      \bottomrule
    \end{tabular}
  \end{threeparttable}
  \normalsize
\end{table}
\vspace*{\fill}

\newpage

\vspace*{\fill}
\begin{table}[!h]
  \begingroup\centering
    \caption{\label{tabS5}Results of fits to \htwo\ MRD profiles from immobilized 
             MbCO at pD~7.2 and 25~$\degc$ with and without 8~bar Xe.$\rsup{a}$}
    \small
    \begin{tabular}{lcc}
      \toprule
      parameter (unit) & $-$Xe & $+$Xe \\
      \midrule
      $\tau_{\mathrm{C},1}$ (\us)    & $5.6 \pm 0.8$   & [5.6]           \\
      $N_1$                          & [6.3]           & $6.32 \pm 0.03$ \\
      $S_1$                          & $0.78 \pm 0.02$ & [0.78]          \\
      $\eta_1$                       & $0.4 \pm 0.2$   & [0.4]           \\
      $\tau_{\mathrm{C},2}$ (ns)     & $127 \pm 17$    & [127]           \\
      $N_2 S_{\mathrm{iso},2}^{\,2}$ & $0.69 \pm 0.05$ & $0.71 \pm 0.2$  \\
      $\tau_{\mathrm{C},3}$ (ns)     & $6.5 \pm 1.3$   & [6.5]           \\
      $N_3 S_{\mathrm{iso},3}^{\,2}$ & $3.8 \pm 0.8$   & $3.6 \pm 0.3$   \\
      $\xi\rsub{H}$                  & $8.1 \pm 0.8$   & $8.3 \pm 0.6$   \\
      $\chi\rsub{red}^{\,2}$         & 3.15            & 2.38            \\
      \bottomrule
    \end{tabular}
  \par\endgroup
  \vspace*{5mm}
  \footnotesize
  $\rsup{a}$ Due to slightly larger data scatter, a constrained fit was 
             performed to the MRD profile in the absence of Xe, with $N_1 = 6.3$ 
             fixed at the value expected from the pD~7.0 fit (Fig.~\ref{fig2}a) 
             and His $\pka$ values (Table~\ref{tabS2}). 
             For the MRD profile in the presence of Xe, the unconstrained 
             parameter values do not differ significantly from those obtained in 
             the absence of Xe, demonstrating that the two data sets are 
             indistinguishable within experimental error.\par
  \normalsize
\end{table}

\vspace*{\fill}

\vspace*{\fill}

\begin{landscape}
\vspace*{\fill}
\begin{table}[!h]
  \centering
  \caption{\label{tabS6}Internal water molecules in crystal structures of horse 
           heart Mb.}
  \small
  \begin{tabular}{lcccclccl}
    \toprule
    \multirow{3}{*}{PDB} & \multirow{3}{*}{form} & \multirow{3}{*}{res.~(\AA)} 
    & \multirow{3}{*}{$T$ (K)} & \multirow{3}{*}{pH} 
    & \multicolumn{4}{c}{internal water characteristics}\\
    \cmidrule(lr){6-9}
    & & & & & site (water no.) & occup. & \textit{B} factor 
    & H-bond acceptor/donor (length in {\AA}) \\
    \midrule
    \multirow{2}{*}{1HRM\cite{Hildebrand1995}} & \multirow{2}{*}{H93Y, met?} 
    & \multirow{2}{*}{1.70} & \multirow{2}{*}{n/a} & \multirow{2}{*}{n/a} 
    & Xe3a (160) & 1.0 & 18.25 & I75 O (2.9), G80 N (2.8), H82 N$^{\updelta 1}$ 
    (2.9), Xe3b (3.4) \\[2pt]
    & & & & & Xe3b (159) & 1.0 & 34.00 & A134 O (2.8), Xe3a (3.4) \\
    \midrule
    \multirow{4}{*}{1WLA\cite{Maurus1997}} & \multirow{4}{*}{met} 
    & \multirow{4}{*}{1.70} & \multirow{4}{*}{n/a} & \multirow{4}{*}{n/a} 
    & DP (156) & 1.0 & 8.26 & Heme Fe, H64 N$^{\upvarepsilon 2}$ \\[2pt]
    & & & & & Xe3a (180) & 1.0 & 10.89 & I75 O (2.9), G80 N (2.7), 
    H82 N$^{\updelta 1}$ (3.0), Xe3b (3.0) \\[2pt]
    & & & & & Xe3b (216) & 1.0 & 38.98 & A134 O (3.1), Xe3a (3.0) \\[2pt]
    & & & & & apical (217)& 1.0 & 36.46 & Q26 N$^{\upvarepsilon 2}$ (2.6), 
    M55 O (2.7), S58 O (3.0) \\
    \midrule
    \multirow{4}{*}{1DWR\cite{Chu2000}} & \multirow{4}{*}{CO} 
    & \multirow{4}{*}{1.45} & \multirow{4}{*}{100} & \multirow{4}{*}{7.5} 
    & Xe3a (2072) & 1.0 & 10.01 & I75 O (2.9), G80 N (2.8), 
    H82 N$^{\updelta 1}$ (2.8), Xe3b (3.0) \\[2pt]
    & & & & & Xe3b (2106) & 1.0 & 18.53 & A134 O (3.0), Xe3a (3.0) \\[2pt]
    & & & & & apical (2059) & 1.0 & 13.91 & Q26 O$^{\upvarepsilon 2}$ (2.7), 
    M55 O (2.8), S58 O (2.6) \\[2pt]
    & & & & & bottom (2078) & 1.0 & 13.00 & Q91 N$^{\upvarepsilon 2}$ (3.2), 
    K145 O (3.0), E148 O$^{\upvarepsilon 2}$ (2.9) \\
    \midrule
    \multirow{5}{*}{2V1K\cite{Hersleth2007}} & \multirow{5}{*}{deoxy} 
    & \multirow{5}{*}{1.25} & \multirow{5}{*}{110} & \multirow{5}{*}{6.8} 
    & DP (2186) & 1.0 & 32.80 & Heme Fe (3.7), H64 N$^{\upvarepsilon 2}$ 
    (3.6) \\[2pt]
    & & & & & Xe3a (2100) & 1.0 & 14.13 & I75 O (2.8), G80 N (2.9), 
    H82 N$^{\updelta 1}$ (2.8), Xe3b (3.0) \\[2pt]
    & & & & & Xe3b (2155) & 1.0 & 20.92 & A134 O (3.0), Xe3a (3.0) \\[2pt]
    & & & & & apical (2038) & 1.0 & 20.87 & Q26 N$^{\upvarepsilon 2}$ (2.6), 
    M55 O (2.8), S58 O (2.7) \\[2pt]
    & & & & & bottom (2172) & 1.0 & 16.89 & Q91 N$^{\upvarepsilon 2}$ (2.9), 
    K145 O (3.0), E148 O$^{\upvarepsilon 1}$ (2.6) \\
    \bottomrule
  \end{tabular}
  \normalsize
\end{table}
\vspace*{\fill}
\end{landscape}

\begin{landscape}
\vspace*{\fill}
\begin{table}[!h]
  \centering
  \caption{\label{tabS7}Internal water molecules in crystal structures of sperm 
           whale Mb.}
  \small
  \begin{tabular}{lcccclccl}
    \toprule
    \multirow{3}{*}{PDB} & \multirow{3}{*}{form} & \multirow{3}{*}{res.~(\AA)} 
    & \multirow{3}{*}{$T$ (K)} & \multirow{3}{*}{pH} 
    & \multicolumn{4}{c}{internal water characteristics}\\
    \cmidrule(lr){6-9}
    & & & & & site (water no.) & occup. & \textit{B} factor 
    & H-bond acceptor/donor (length in {\AA}) \\
    \midrule
    \multirow{4}{*}{1BZR\cite{Kachalova1999}} & \multirow{4}{*}{CO} 
    & \multirow{4}{*}{1.15} & \multirow{4}{*}{287} & \multirow{4}{*}{5.9} 
    & Xe3a (340) & 1.0 & 18.25 & I75 O (2.8), G80 N (2.9), 
    H82 N$^{\updelta 1}$ (2.8), Xe3b (3.1) \\[2pt]
    & & & & & Xe3b (403) & 1.0 & 31.34 & A134 O (3.1), Xe3a (3.1) \\[2pt]
    & & & & & apical (304) & 0.65 & 18.07 & Q26 N$^{\upvarepsilon 2}$ (2.9), 
    M55 O (2.7), S58 O (2.7) \\[2pt]
    & & & & & W1 (382) & 1.0 & 26.32 & A22 O (2.7), K62 O (2.6) \\
    \midrule
    \multirow{5}{*}{1A6G\cite{Vojtechovsky1999}} & \multirow{5}{*}{CO} 
    & \multirow{5}{*}{1.15} & \multirow{5}{*}{100} & \multirow{5}{*}{6.0} 
    & Xe3a (1006) & 0.86 & 11.34 & I75 O (2.8), G80 N (2.9), 
    H82 N$^{\updelta 1}$ (2.8), Xe3b (3.0) \\[2pt]
    & & & & & Xe3b (1056) & 0.88 & 17.29 & A134 O (2.9), Xe3a (3.0) \\[2pt]
    & & & & & apical (1053) & 1.0 & 14.96 & Q26 N$^{\upvarepsilon 2}$ 
    (2.6--2.9), M55 O (2.9), S58 O (2.8) \\[2pt]
    & & & & & W1 (1011) & 0.86 & 16.32 & A22 O (2.6), K62 O (2.7) \\[2pt]
    & & & & & W2 (1154) & 0.71 & 30.44 & A127 O (2.7) \\
    \midrule
    \multirow{6}{*}{1BZ6\cite{Kachalova1999}} & \multirow{6}{*}{met} & \multirow{6}{*}{1.20} & \multirow{6}{*}{287} & \multirow{6}{*}{6.0} & DP (389) 
    & 1.0 & 8.86 & Heme Fe (2.2), H64 N$^{\upvarepsilon 2}$ (2.7) \\[2pt]
    & & & & & Xe3a (352) & 1.0 & 12.75 & I75 O (2.8), G80 N (2.9), 
    H82 N$^{\updelta 1}$ (2.8), Xe3b (3.1) \\[2pt]
    & & & & & Xe3b (417) & 1.0 & 34.98 & A134 O (3.1), Xe3a (3.1) \\[2pt]
    & & & & & apical (509) & 0.30 & 11.87 & Q26 O$^{\upvarepsilon 1}$ (2.9), 
    M55 O (2.8), S58 O (2.8) \\[2pt]
    & & & & & W1 (393) & 1.0 & 19.83 & A22 O (2.7), K62 O (2.7) \\[2pt]
    & & & & & W2 (500) & 0.50 & 24.49 & A127 O (3.0) \\
    \midrule
    \multirow{6}{*}{1A6K\cite{Vojtechovsky1999}} & \multirow{6}{*}{met} 
    & \multirow{6}{*}{1.10} & \multirow{6}{*}{90} & \multirow{6}{*}{7.0} 
    & DP (1001) & 1.0 & 9.70 & Heme Fe (2.1), H64 N$^{\upvarepsilon 2}$ 
    (2.7) \\[2pt]
    & & & & & Xe3a (1006) & 0.93 & 10.64 & I75 O (2.8), G80 N (2.8), 
    H82 N$^{\updelta 1}$ (2.7), Xe3b (3.0) \\[2pt]
    & & & & & Xe3b (1056) & 0.83 & 14.77 & A134 O (2.9), Xe3a (3.0) \\[2pt]
    & & & & & apical (1053) & 1.0 & 11.55 & Q26 N$^{\upvarepsilon 2}$ 
    (2.6--3.0), M55 O (2.9), S58 O (2.8) \\[2pt]
    & & & & & W1 (1011) & 0.93 & 13.02 & A22 O (2.7), K62 O (2.7) \\[2pt]
    & & & & & W2 (1154) & 0.70 & 23.42 & A127 O (2.9) \\
    \bottomrule
  \end{tabular}
  \normalsize
\end{table}
\vspace*{\fill}
\end{landscape}

\vspace*{\fill}
\vspace*{\fill}

\end{document}